\documentclass[aps, prd, twocolumn, lengthcheck, superscriptaddress, 
nofootinbib]{revtex4-1}

\usepackage{epsfig}
\usepackage[usenames]{color}
\usepackage{graphicx}
\usepackage{amsmath}
\usepackage{multirow}
\usepackage{epstopdf}

\def\bi{\bibitem}

\def\la{\langle}\def\ra{\rangle}
\def\be{\begin{eqnarray}}\def\ee{\end{eqnarray}}
\def\lsim{\mathrel{\rlap{\lower3pt\hbox{\hskip1pt$\sim$}}
     \raise1pt\hbox{$<$}}} 
\def\gsim{\mathrel{\rlap{\lower3pt\hbox{\hskip1pt$\sim$}}
     \raise1pt\hbox{$>$}}} 
\def\del{\partial}

\def\GDS{{\it GDS}}

\allowdisplaybreaks

\begin{document}

\title{Manifestation of Hidden Symmetries in Baryonic Matter: \\ 
From Finite Nuclei to  Neutron Stars}

\author{Mannque Rho}
\email{mannque.rho@ipht.fr}
\affiliation{Universit\'e Paris-Saclay, CNRS, CEA, Institut de Physique Th\'eorique, 91191, Gif-sur-Yvette, France }

\author{Yong-Liang Ma}
\email{ylma@ucas.ac.cn}
\affiliation{School of Fundamental Physics and Mathematical Sciences,
Hangzhou Institute for Advanced Study, UCAS, Hangzhou, 310024, China}
\affiliation{International Center for Theoretical Physics Asia-Pacific, Beijing/Hangzhou, China }

\date{\today}

\begin{abstract}
The hadron-quark/gluon duality formulated in terms of a topology change at a density $n\gsim 2n_0$ $n_0\simeq 0.16$fm$^{-3}$ is found to describe the core of massive compact stars in terms of  quasiparticles of fractional baryon charges, behaving neither like pure baryons nor like deconfined quarks. This can be considered as the Cheshire-Cat mechanism~\cite{CC}  for the hadron-quark continuity arrived at bottom-up from skyrmions that is equivalent to  the ``MIT-bag"-to-skyrmion  continuity arrived at top-down from quarks/gluons.  Hidden symmetries, both local gauge and pseudo-conformal (or broken scale),  emerge and give rise both to  the long-standing ``effective $g_A^\ast\approx 1$" in nuclear Gamow-Teller transitions at $\lsim n_0$ and to the pseudo-conformal sound velocity $v_{pcs}^2/c^2\approx 1/3$ at $\gsim 3n_0$. It is suggested that what has been referred to, since a long time, as  ``quenched $g_A$" in light nuclei  reflects what leads to the dilaton-limit $g_A^{\rm DL}=1$ at near the (putative) infrared fixed point of scale invariance. These properties are confronted with the recent observations in Gamow-Teller transitions and in astrophysical observations.
\end{abstract}

\maketitle

\section{Introduction}
While the structure of nuclear matter at density $n=n_0\approx 0.16$ fm$^{-3}$ is fairly well understood, the phase structure of the strong interactions at high densities,  investigated for several decades,  still remains largely uncharted. Recent precision measurements of massive $\gsim 2 M_\odot$ neutron stars and detection of gravitational waves from  star mergers provide indirect information of nuclear matter at low temperature and at high density, say,  up to  ten times the saturation density $n_0$. So far, such phenomena can be accessed by neither terrestrial experiments nor lattice simulation, the only model-independent tool known in strong-interaction physics.

The study of dense matter in the literature has largely relied on either largely phenomenological approaches anchored on density functionals or effective field theoretical models implemented with certain QCD symmetries, constructed in terms of set of relevant degrees of freedom appropriate for the cutoff chosen for the effective field theory (EFT),  such as  baryons and  pions, and  with~\cite{baymetal,quarkyonic} or without~\cite{HRW,DHW2021}  hybridization with quarks, including other massive degrees of freedom.  
The astrophysical observations indicate that the density probed in the interior of neutron stars could be as high as $\sim 10$ times the normal nuclear matter density $n_0$  and immediately raise the question as to what the interior of the star could consist of, say, baryons and/or quarks  and a combination thereof. Asymptotic freedom of QCD implies that at some superhigh  density, the matter could very well be populated by deconfined quarks~\cite{collins-perry}. But the density of the interior of stars is far from the asymptotic and hence perturbative QCD cannot be reliable there. Now the question that immediately arises is whether and how the relevant degrees of freedom of QCD, namely, the gluons and quarks intervene in the high density regime relevant to the interior of massive compact stars. Given the apparently different degrees of freedom from hadrons at low density and the quark-gluon degrees of freedom presumably figuring at high density, does addressing massive star physics  involve one or more phase transitions?

In this review, we describe a conceptually novel approach going beyond the standard chiral EFT (denoted as sChEFT) to higher densities $n\gg n_0$ formulated by us since some years and exploit it to  not only post-dict -- with success -- the experimentally studied properties of baryonic matter at density near $n_0$ but also predict high density properties of massive compact stars. Given that at low densities, the relevant degrees of freedom are hadrons while at some high densities, they must be quarks and gluons of QCD, it seems reasonable to presume that the proper approach would involve going from hadronic theory to quark-gluon theory. At low densities it is clear that the appropriate theory is an EFT anchored on chiral symmetry, i.e., $\chi$ EFT, but what is the appropriate theory at high density which cannot be directly accessed by QCD proper? It must be necessarily an EFT but not knowing what the UV completion of an EFT is, there is no obvious way to formulate a semi-microscopic theory involving quarks and gluons. Currently resorted to are the class of linear sigma models such as Nambu-Jona-Lasinio-type theories  or  bag models with or without coupling to hadrons such as pions, vector mesons etc. with arbitrary constants phenomenologically adjusted with no link to QCD. The strategy there is then to hybrid to certain extent arbitrarily a hadronic description to the putative quark/gluonic description at a suitable density regime where the changeover of degrees of freedom is presumed to take place.  

 The merit of the approach that we rely on is that we will have a single unified, admittedly perhaps {\it oversimplified}, effective Lagrangian formulated in a way that encompasses from low density to high density, involving only manifestly ``macroscopic" degrees of freedom, but capturing the continuity to ``microscopic" quarks-gluon degrees of freedom. How to effectuate the change of degrees of freedom will be formulated in terms of a possible topology change at a density denoted $n_{1/2}$ encoded in the behavior of the parameters of the EFT Lagrangian as one moves from below  to above the changeover density $n_{1/2}$.

\section{Hidden Symmetries and New Degrees of Freedom}\label{hiddensymmetries}
We start by specifying the degrees  of freedom that figure in our formulation of the theory.  For consistency with the scale symmetry discussed below, we need the number of flavors to be 3 with nearly ``massless" u(p), d(own) and s(trange) quarks. We will however confine ourselves to the u and d quarks in most of what follows. In what way, the s flavor figures will be explained below. 

Since we will be dealing with an effective field theory (EFT), the degrees of freedom figuring in the theory will depend on the energy/momentum (or length) scale involved. Specifically in terms of density at zero temperature with which we will be concerned here,  if we are interested in nuclear interactions in the vicinity of nuclear equilibrium density, say, $n\lsim 2n_0$, then the relevant energy scale can be slightly less than the mass of the lowest-lying vector-meson, $\rho$,  i.e., so  $\Lambda\sim (400-500)$ Mev. The relevant degrees of freedom will then be the pions $\pi$ in addition to the nucleons $N^T=(n p)$. The nucleon has mass higher than the scale $\Lambda$, but must figure for nuclear physics either directly or as a skrymion. This implies that it makes sense only if nuclear dynamics involved are ``soft" relative to the cutoff,  in the sense the pion mass $m_\pi$ is soft.  The EFT that comprises of pions and nucleons, suitably anchored on chiral symmetry of QCD,  is the {\it standard} chiral EFT (denoted sChEFT) which has been established, when formulated in {\it ab inito} approach exploiting powerful numerical techniques,  to be fairly successful in nuclear physics at low energy and at density near $n_0$. The well-defined chiral power expansion with systematic renormalization of higher-order terms allows one to account for scalar and vector-channel excitations more massive than the pion. This sChEFT, currently feasible in practice up to N$^n$LO for $n \lsim 4$ must, however, break down as density approaches the scale comparable to the vector-meson and scalar meson mass $\lsim 700$ MeV. Where this breakdown takes place is not known precisely but it is considered to be very likely to be of $\gsim 2n_0$ involving  the possible change of degrees of freedom from hadrons to {\it effective} quarks/gluons. In our approach this changeover will be identified with ``hadron-quark continuity" {\it without deconfinement}.

The key point in the approach is that in going above $n_0$ toward compact-star densities $\sim (5-7)n_0$, two hidden symmetries play a crucial role in capturing the hadron-quark continuity.  The symmetries involved are hidden local symmetry (HLS for short) for $V_\mu$ and hidden scale symmetry for the dilaton that we denote by $\chi$ corresponding  to $f_0(500)$\footnote{In the literature, it is denoted as $\sigma$ which transforms nonlinearly under scale transformation. We use $\chi$ that transforms linearly as will be explained.}. This means that the appropriate cutoff of the EFT is set above the lowest vector and scalar mass. In some cases, a higher tower of vector mesons could figure as in holographic models but we will have them integrated out in this review except when they are needed as will be discussed later. There are also other more subtle symmetries such as parity-doubling, pseudogap symmetry etc. which will figure associated with the two main hidden symmetries. They will be explained when needed.

\subsection{Hidden local symmetry (HLS)}
We first address in this section how to bring in the hidden local symmetry into the chiral Lagrangian that is the basis for sChEFT.  The hidden scale symmetry will be treated in the next section.

In the chiral $SU(2)_L\times SU(2)_R$ symmetry spontaneously broken to $SU(2)_{V=L+R}$, there is a  redundancy in the chiral field $U$
\be
U = e^{2i\pi/f_\pi}=\xi_L^\dagger\xi_R &=& \xi_L^\dagger h(x)h(x)^\dagger \xi_R, \\  
h(x) &\in&  SU(2)_{V}.\nonumber
\ee 
Gauging this redundancy in a particular way picked by ~\cite{yamawaki,HY:PR} leads to what is referred to in the literature as ``hidden local symmetry (HLS)"\footnote{Extensive references particularly  relevant to this review are found in \cite{HY:PR}.}. ``Gauging" the redundancy by itself is of course empty but endowing it with a kinetic energy term makes the vector field propagate and hence become gauge field. The basic assumption is that the local (gauge) symmetry is dynamically generated. As such there is nothing that implies that that symmetry is intrinsic in QCD, and hence  could very well be a gauge symmetry emergent from the strong dynamics as in strongly correlated condensed systems. According to Suzuki~\cite{suzuki}, a gauge-invariant local field theory written in {\it matter fields alone} {\it inevitably} gives rise to composite gauge bosons (CGB).\footnote{The argument involves subtle notions of global, local and gauge symmetries juxtaposed with physical symmetries anchored on Noether's theorem~\cite{noether}.} This theorem would then imply that the HLS as constructed in the way stated above must, in the chiral limit (i.e., $m_u=m_d=0$),   have what is called in \cite{HY:PR} as the ``vector manifestation" (VM) fixed point at which the vector meson mass goes to zero.  

There are two important observations to make here -- to be elaborated in greater details later -- on the notion of the vector mesons as composite gauge fields. 

The first is the role of $V_\mu$ channels in nuclear dynamics which is encoded in sChEFT at high orders in chiral perturbation series. Now the HLS Lagrangian is gauge-equivalent to non-linear sigma model which is the basis of sChEFT. This means that \`a la Suzuki theorem, sChEFT, correctly treated, should give rise to the VM fixed-point structure. In other words, there must exist at some high density (or temperature) a phase where the $m_V=0$.\footnote{Such a phase should be visible at high temperature in heavy-ion dilepton processes. If Nature confirms the absence of such a phase it would then rule out the composite gauge structure of the vector mesons. We will argue later that up to date a possibly crucial phase structure associated with an $\eta^\prime$ singularity -- a topological object involving Chern-Simons topological field -- has been ignored at near chiral phase transition.}

The second observation is that the HLS fields described as CGB fields could become ``Seiberg-dual"~\cite{Komargodski} at high density (or temperature)  to the gluons of QCD.  This would mean that HLS  is actually encoded -- but hidden -- in QCD proper and emerges un-hidden at high density (or temperature).  How this feature manifests in experiments is quite subtle and not obvious because of the topological structure that involves an $\eta^\prime$ singularity at high density, possibly associated with quantum Hall droplet baryons~\cite{QHdroplet,CCP,karasik,karasik2,kanetal,D}.  

\subsection{Hidden scale symmetry}
\subsubsection{``Genuine dilaton scheme"}
\begin{figure}[htbp]
\begin{center}
\includegraphics[width=0.5\textwidth]{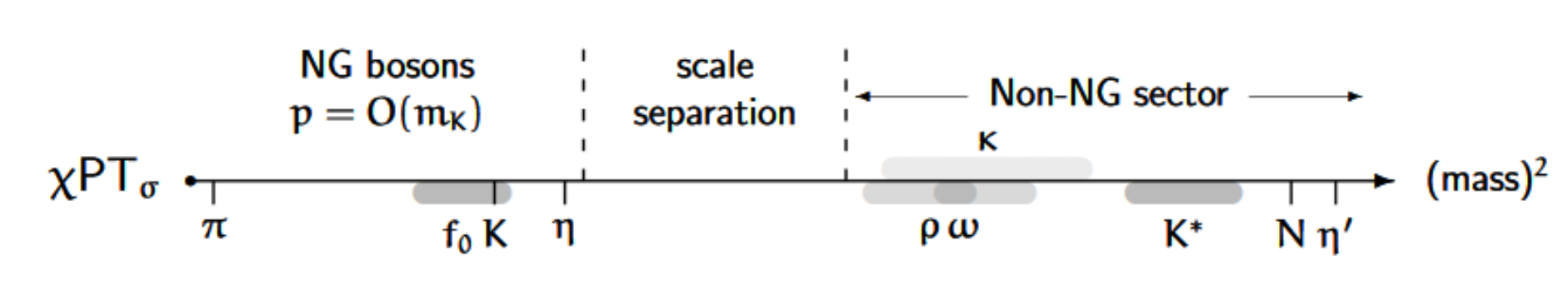}
\end{center}
\vskip -.5cm
\caption{Scale separation for chiral-scale perturbation theory (denoted $\chi$PT$_\sigma$) in \GDS\  copied from \cite{GDS}}\label{chipt}
\end{figure}

In QCD, scale symmetry is broken by quantum (a.k.a. trace) anomaly and quark masses\footnote{In what follows, quark masses will not be explicitly shown in the discussions although they are implicitly  included in numerical results given in the review. How the quark masses figure is given in the ``genuine dilaton scheme."}.  In our approach we adopt the ``genuine dilaton" scheme~\cite{GDS}  for scale symmetry. In this scheme the scalar $f_0(500)$ is considered to be as light as the kaons $K$, and classed as a scalar pseudo-Nambu-Goldstone boson (of spontaneously broken scale symmetry) -- called ``dilaton" from here on -- put on par with the kaons in the pseudo-scalar pseudo-NG bosons. The genuine dilaton scheme  requires taking the flavor symmetry $SU(3)$. Unless otherwise noted, however, we will continue our discussion with 2 flavors. Later $\eta^\prime$ will enter accounting for the 3rd flavor. Together with the vector mesons $V_\mu$, the mass scales involved are given in Fig.~\ref{chipt} copied from \cite{GDS}. Combined with the three-flavor chiral symmetry, the dilaton gives rise to ``chiral-scale symmetry" captured at low energy by chiral-scale perturbation theory $\chi$PT$_\sigma$.  As indicated, in the matter-free vacuum, there is a net scale separation between the NG sector and the massive sector denoted as non-NG sector.  But in nuclear medium at high density, the scale separation between the NG sector and the vector mesons in the NG sector becomes blurred. Also the $\eta^\prime$ is most likely to join the NG sector at high density. These features will be elaborated in what follows.

We should mention that this genuine dilaton scheme ({\GDS} for short) -- which has been extended to BSM (beyond the Standard Model) involving dilatonic Higgs physics -- seems to be fundamentally different from the currently popular scenario involving the conformal window for $N_f\sim 8$ or greater. We shall not go into the presently controversial issues involved in the matter. As we will develop, however, this scheme -- which will be associated with ``emergent symmetries" from nuclear correlations -- is surprisingly, though not rigorously provable, consistent with nuclear phenomena.

Our main interest is in combining this \GDS\ with HLS for treating nuclear interactions in the density regime encompassing from finite nuclei to dense compact-star matter density. It will be shown that this \GDS --  with its associated {\it soft (NG boson) theorems} -- is particularly predictive and strikingly consistent with what takes place in nuclear interactions both at low and high densities . 

The idea is to extend the chiral perturbation strategy established in HLS~\cite{HY:PR} to a combined ``chiral-scale symmetry" scheme that incorporates the power counting due to scale symmetry.  In \cite{GDS}, neither vector bosons nor baryons were explicitly addressed. The scheme combining these degrees of freedom essential for nuclear physics  --  that we dub $\psi d$HLS  with $\psi$ standing for baryon and $d$ for dilaton -- was first formulated in \cite{LMR}.\footnote{In its present form, it looks much too complicated with exploding parameters although the basic idea involved is extremely simple. It could surely be made much simpler and more readily manageable for applications in nuclear physics, a work for the future.}

\begin{figure}[htbp]
\begin{center}
\includegraphics[width=0.4\textwidth]{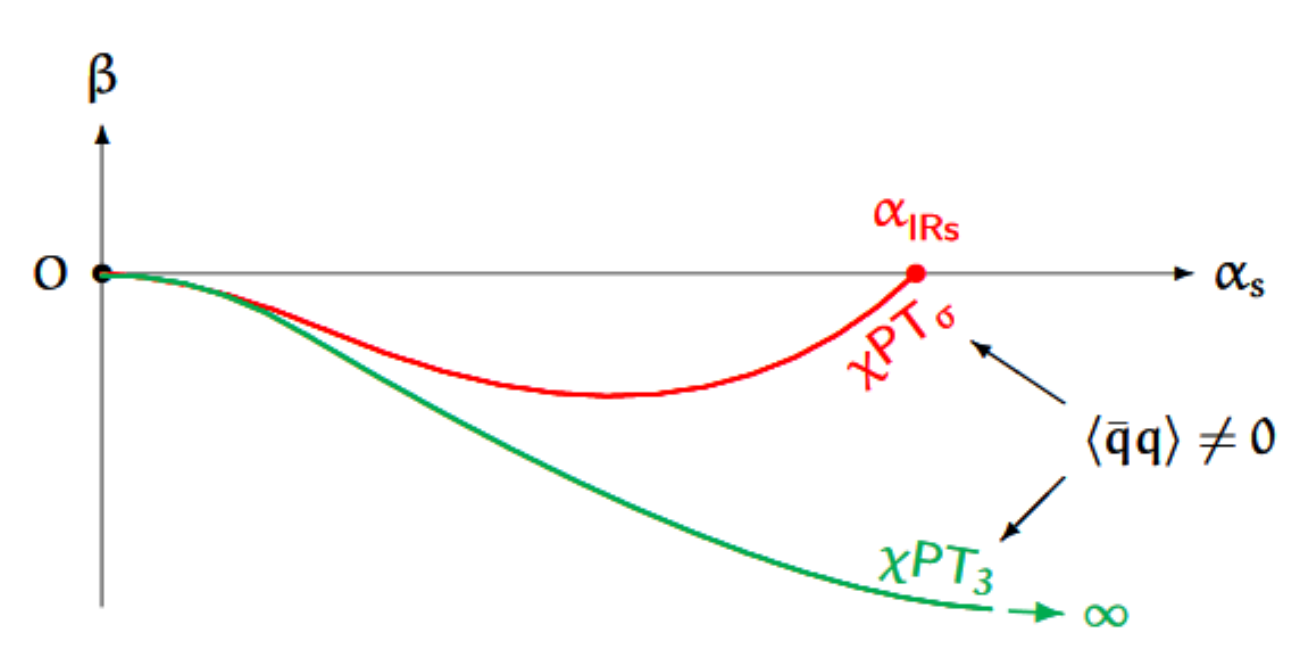}
\end{center}
\vskip -.5cm
\caption{$\beta$ function flows for the $N_f=3$ QCD ($\chi$PT$_3$) and for the genuine dilaton flow ($\chi$PT$_\sigma$): Copied from \cite{GDS}.}\label{beta}
\end{figure}

What is distinctive of this \GDS\ is that in the chiral limit\footnote{Unless otherwise stated, we will always be dealing with the chiral limit.}, there is an infrared (IR) fixed point $\alpha_{\rm IRs}$ -- at which the QCD $\beta$ function vanishes   -- as indicated  in Fig.~\ref{beta} in the $\chi$PT$_\sigma$ flow. It is contrasted in the figure with the $\beta$ function flow for $N_f=3$ denoted  $\chi$PT$_3$. The IR fixed point is characterized by that conformal invariance is realized in NG mode with $f_\pi\sim f_\sigma \neq 0$ with massive matter fields accommodated.  The striking difference of this scheme from that based on conformal window adopted by many authors in the field of BSM (see, e.g., \cite{appelquist} and references given therein)  where the IR fixed point involved, that is,  conformal symmetry,  is realized  in  Wigner mode with $f_\pi=f_\sigma=0$.
\subsubsection{Chiral-scale symmetry}\label{CSS}
We briefly explain  how to combine scale symmetry to hidden local gauge symmetry explained above so as to perform chiral-scale perturbation expansion. Details can be found in \cite{GDS,LMR,MR-review}. For this, it is convenient to use the ``conformal compensator field"  $\chi$ related to the dilaton field $\sigma$ by
\be
\chi=f_\chi e^{\sigma/f_\chi}.
\ee
In this form, $\chi$ -- with mass dimension 1 -- transforms linearly under scale transformation whereas $\sigma$ transforms nonlinearly.  From hereon when we refer to dilaton, we shall mean $\chi$ instead of $\sigma$.

Given that HLS is generated dynamically from the non-linear sigma model, the chiral-scale power counting of HLS Lagrangian is straightforward. What is needed is the power counting in scale symmetry.  Among various power counting schemes that have been made in the literature (e.g., \cite{GS-2020} in the large $N_f$ context), we adopt the \GDS. The ``small" expansion parameter in scale symmetry, apart from the power of derivatives $O(\del^2)\sim O(p^2)$,  is the deviation of the $\beta$ function from the IR fixed point $\beta (\alpha_{\rm IRs})$. Expanding the $\beta$ function near the IR fixed point, $\beta (\alpha_s)=\epsilon\beta^\prime +O(\epsilon^2)$ where $\epsilon=\alpha_s-\alpha_{\rm IRs}$ and $\beta^\prime=\del\beta (\alpha_s)/\del\alpha_s|_{\alpha_s=\alpha_{\rm IR}}$ is  the anomalous dimension of the gluon stress tensor ${\rm tr} G_{\mu\nu}^2$. It is clear from Fig.~\ref{beta} that $\beta^\prime >0$, so the chiral-scale power counting is 
\be
\epsilon \sim  O(p^2).
\ee
In accordance of chiral-scale symmetry,  the dilaton mass scales as the pion mass does
\be
m_\chi\sim O(\epsilon)\sim O(p^2).
\ee

Now the chiral-scale HLS~\cite{LMR} (denoted $d$HLS) can be organized in the same way as with nonlinear sigma model without the vector fields~ \cite{GDS}. Take for example the leading chiral order (LO)  HLS with scale dimension $d_s=2$ which is made scale-invariant by multiplying by  $\big(\frac{\chi}{f_\chi}\big)^2$. Denote it ${\cal L}_{\rm HLS:LO}$. The \GDS\  is to  
take into account the anomalous dimension $\beta^\prime$ calculated by Callan-Symansik renormalization group equation to incorporate the scale anomaly effect 
\be
{\cal L}_{\rm LO} = q(c,\beta^\prime) {\cal L}_{d\rm HLS:LO}\label{LO}
\ee
where
\be
q(c,\beta^\prime)=\Big(c+(1-c)\big(\frac{\chi}{f_\chi}\big)^{\beta^\prime}\Big).
\ee
Here $c$ is a  constant not given by symmetry alone. Note that the anomalous dimension effect is present only when $c\neq 1$. The first term has scale dimension $d_s=4$, hence scale-invariant but the second term has $d>4$ and hence breaks scale invariance. One can see from Fig.~\ref{beta} that $\beta^\prime$ cannot be zero in this scheme. 

All chiral-scale leading order terms can be given both $d_s=4$ and $d>4$ terms with different coefficients $c_J$ for different $J$. How to go to chiral N$^n$LO terms for $n>2$ is not difficult to formulate and this is worked out in \cite{LMR} including couplings to baryon fields. 

\subsubsection{Soft Theorems}\label{softtheorems}
It should be stressed that {\it a striking feature of this \GDS  -- which will play a key role in what's discussed below -- is that approaching the IR fixed point ``soft theorems" apply to both scalar and pseudo-scalar NG bosons.}  Both play a very powerful role to the developments  made below in nuclear processes. In the \GDS, the soft theorems are closely parallel between chiral symmetry and scale symmetry. Among various soft theorems, the Goldberger-Treiman relation for the pion $ f_\pi g_{\pi NN} = g_A m_N$ has the counterpart for the dilaton\footnote{We announce in advance that the pionic and dilatonic Goldberger-Treiman relations applied to nuclear systems will be found to lead to $g_A=1$ in dense matter at what will be called ``dilaton-limit fixed point."}.
\be
f_\chi g_{\chi NN} = m_N \label{GTdilaton}
\ee
and the dilaton counterpart for the Gell-Mann-Oakes-Renner (GMOR) relation for the pion $m_\pi^2 f_\pi^2 = - \bar{m}\la\bar{q}q\ra_0$ is
\be
m_\chi^2 f_\chi^2 = \frac{\epsilon \beta^\prime}{\alpha_{IR}} \la G_{\mu\nu}^2\ra_0 \ .\label{GMORdilaton}
\ee  
There should of course be nonzero quark mass term in this expression for the dilaton which is ignored in the chiral limit. Note that it is $\beta^\prime\epsilon$ that plays here the role of ``explicit" symmetry breaking due to quantum anomaly even in the absence of the quark masses. 

In the matter-free vacuum, these soft theorems  are verified to hold very well for the pion, e.g., to high orders in chiral perturbation theory and also on lattice QCD.  They are also verified to hold fairly well in chiral perturbation theory in nuclear matter (see, e.g., \cite{oller} for the GMOR relation). Disappointingly there are up-to-date no higher-order calculations  that give support to soft-dilaton theorems  both in the vacuum and in medium although arguments are given in \cite{GDS} that suggest that they will work at least as well as soft-kaon theorems in the vacuum. It will be shown however that combined with soft-pion theorems, soft-dilaton theorems do resolve a long-standing puzzle of quenched $g_A$ in nuclear Gamow-Teller transitions~\cite{gA}.  This, together with what takes place in highly dense compact-star matter, will be the key contribution of this review.
\subsubsection{Chiral-scale HLS and ``LOSS" approximation}\label{LOSSA}
In applying the chiral-scale HLS Lagrangian described above to nuclear processes at normal nuclear systems at $n\sim n_0$ and highly compressed matter at $n\sim (5-7) n_0$ relevant to compact stars, we will employ two versions of the same effective Lagrangian, one purely bosonic incorporating the pseudo-scalar NG bosons $\pi\in SU(3)$, the dilaton $\chi$ and the vectors $V$ in consistency with the symmetries concerned.  We are denoting  the resulting scale-symmetric HLS Lagrangian as $d$HLS with $d$ standing for dilaton. To do nuclear physics with it, baryons will be generated as skyrmions with the $d$HLS Lagrangian.
And the second is to introduce baryon fields explicitly into chiral-scale HLS, coupled scale-hidden-local symmetrically to $\pi$ and $\chi$. This Lagrangian will be denoted as $\psi d$HLS with $\psi$  standing for baryons.

At this point we should underline an approximation  dubbed ``leading-order scale symmetry (LOSS)" approximation\footnote{This terminology may very well be a misnomer.} which has figured in applications to compact-star properties in our approach~\cite{MR-review}. The LOSS approximation would be valid if $\beta^\prime \ll 1$ in which case the chiral-scale HLS can be written in terms of scale-invariant $d_s=4$ terms.  All scale-symmetry explicit breaking  can then be put in the Coleman-Weinberg-type dilaton potential. However we will see that in some cases $\beta^\prime \ll 1$ is highly unlikely. It will be shown that  $d>4$ terms can contribute to physical observables with $\beta^\prime {\not\ll} 1$, particularly in  nuclear Gamow-Teller transitions.

\section{Hadron-Quark Continuity}

Given the two Lagrangians $d$HLS and $\psi d$HLS, how does one go about doing nuclear many-body problem ?

While in principle feasible, there is, up to date, no simple way to systematically and reliably formulate nuclear many-body dynamics in terms of skyrmions with $d$HLS.  We will therefore resort to $\psi d$HLS. With the $\psi d$HLS Lagrangian whose parameters are suitably fixed in medium matched to QCD correlators~\cite{HY:PR,MR-review}, we have formulated a Wilsonian renormalization-group (RG)-type approach to arrive at Landau(-Migdal) Fermi-liquid theory along the line developed in condensed matter physics~\cite{shankar}. The mean field approximation has been identified with Landau Fermi-liquid fixed point (FLFP) theory,  valid in the limit $1/\bar{N}\to 0$ where $\bar{N}=k_F/(\Lambda-k_F)$ (with $\Lambda$ the cutoff on top of the Fermi sea)~\cite{MR91,FR}. This approach can be  taken as a generalization of the energy-density functional theory familiar in nuclear physics~\cite{MR-review}. One can go beyond the FLFP in  $V_{lowk}$-RG by accounting for selected $1/\bar{N}$ corrections as we have done in numerical calculations.

As will be briefly summarized below (see \cite{MR-review} for details), the $\psi d$HLS applied to nuclear matter --- call it $Gn$EFT --- is found to describe nuclear matter at $n\sim n_0$ as well as the currently successful standard $\chi$EFT to typically N$^n$LO for $n\leq 4$. There is  however a good reason to believe, although not rigorously proven, that the standard ChEFT with the Fermi momentum $k_F$ taken as a small expansion parameter must inevitably breakdown at high densities relevant to massive compact stars, say, $\gsim 3 n_0$.  Quark degrees of freedom in one form or other could be a natural candidate for the breakdown mechanism.  Our approach has the merit that  where and how this breakdown occurs can can be inferred from the skyrmion approach with the $d$HLS Lagrangian. The strategy is to exploit topology encoded in the dilaton-HLS Lagrangian to access the putative {\it hadron-to-quark continuity} in QCD. The way we approach this is distinctively novel in the field and could surely be further refined but the results obtained thus far are promising as we will show. There are certain predictions that have not been made in other approaches. 
\subsection{Topology change}
We shall eschew going into details that are abundantly available elsewhere~\cite{MR-review} and focus on the key aspects that are new without trying to be precise.

Given that topology is in the pion field in skyrmion approaches to compressed baryonic matter, the original Skyrme model (with  the Skyrme quartic term) -- implemented with the dilaton scalar --  suffices for the purpose at hand. To differentiate it from other skyrmions such as  with the $d$HLS Lagrangian,  used in our approach , or with holographic QCD models, we will specifically refer to the Skyrme model implemented with the dilaton as skyrmion$_{d\pi}$. Otherwise ``skyrmion" will refer to the generic topological  baryon with various different Lagrangians.

The skyrmion approach to baryonic matter, while still far too daunting to handle nuclear dynamics directly and systematically\footnote{Even nucleon-nucleon potentials in the Skyrme model with the pion are still to be worked out satisfactorily  after many decades since  the model was revived in early 1980's~\cite{NNpot}}, can however provide  valuable and robust information on the possible topological structure involved in going beyond the normal nuclear matter density regime. One can exploit its topological structure in dense matter by putting skyrmions on crystal lattice. By reducing the lattice size $L$, one can compress the skyrmion matter. It has been found that at some density denoted $n_{1/2}$ (corresponding to $L_{1/2}$ in the crystal)) the skyrmion matter simulated on, say, FCC lattice is found to turn to a half-skyrmion matter.  This transition, being topological, is considered to be robust. At what density this transition takes place depends on what's in the Lagrangian and hence cannot be determined by theory.  We will determine it by phenomenology below. 

Now the most important feature of this transition is that the chiral condensate $\Sigma\equiv \la\bar{q}q\ra$, the order parameter of chiral symmetry,  which is nonzero in the skyrmion matter for $n < n_{1/2}$, goes, when space averaged, goes to zero ($\bar{\Sigma}\to 0$) for $n\geq n_{1/2}$ while nonzero locally supporting chiral wave. This implies that chiral symmetry remains spontaneously broken with nonzero pion decay constant $f_\pi\neq 0$ in $n\gsim n_{1/2}$ although $\bar{\Sigma}=0$. This resembles the pseudogap structure in superconductivity  where the order parameter is zero but the gap is not. In fact there are other phenomena in strongly correlated condensed matter systems with such pseudogap structure. 

The skyrmion-half-skyrmion transition in dense matter has an analog in condensed matter in (2+1) dimensions, for instance, the transition from the magnetic N\'eel ground state to the VBS (valence bond solid) quantum paramagnet phase~\cite{DQCP}.  There the half-skyrmions intervening in the phase transitions are deconfined  with no local order parameters.  What takes place in dense skyrmion matter seems however quite different because the half-skyrmions remain confined by monopoles~\cite{cho}. This suggests that the confined half-skyrmion complex be treated as a local baryon number-1 field with its parameters, such as the mass, coupling constants etc. drastically modified reflecting the topology change from the vacuum. This observation will be seen to play a key role in ``embedding" the topological inputs in  $Gn$EFT. The topology change from below to above $n_{1/2}$ is {\it to capture in terms of hadronic variables the hadron-quark change-over involving no phase transition, i.e., no deconfinement}.

Our key strategy is to incorporate these topology effects embedded in $d$HLS into the parameters of the $\psi d$bHLS  in formulating $Gn$EFT. 
\subsection{From nuclear symmetry energy to HLS gauge coupling}
One of the most remarkable impacts of the topology change at $n_{1/2}$ is its effect on the nuclear symmetry energy $E_{sym}$ in the equation of state (EoS) of baryonic matter and on the HLS gauge coupling constant, specially  $g_\rho$,  as the vacuum is modified by density. We can work with the skyrmions with $d$HLS but the argument can be made with the skyrmion$_{d\pi}$.

\begin{figure}[htbp]
\begin{center}
\includegraphics[width=0.4\textwidth]{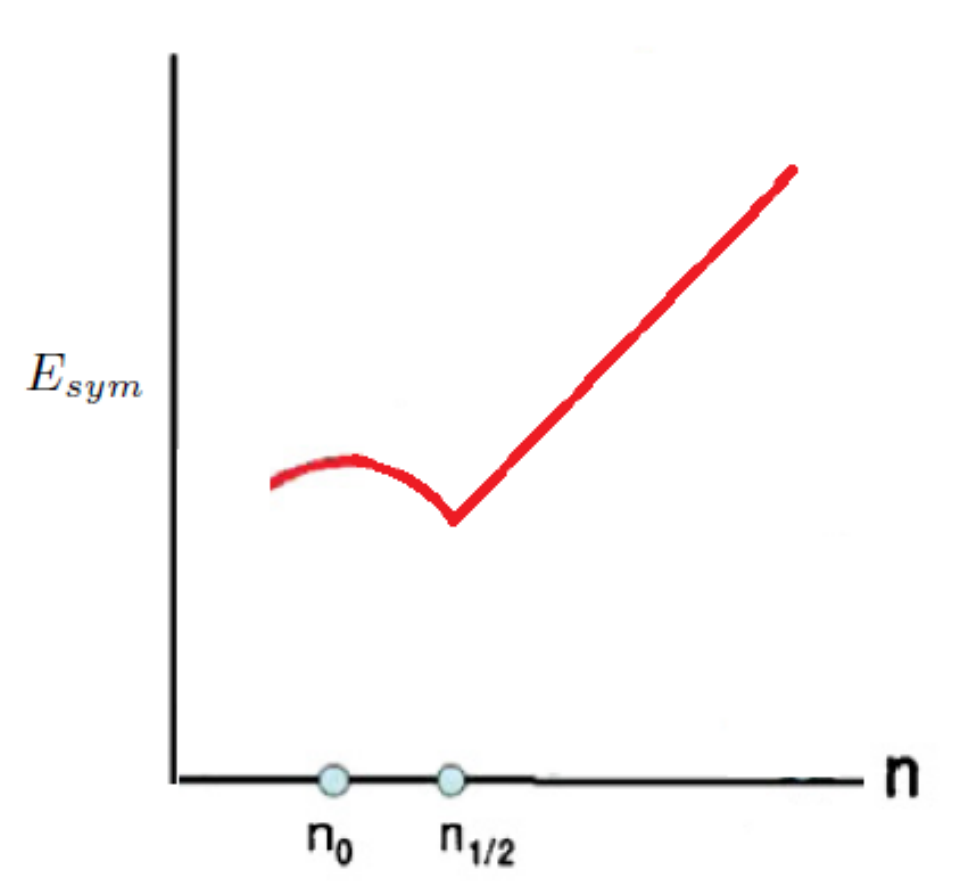}
\end{center}
\vskip -.5cm
\caption{Schematic cusp structure of $E_{sym}$ at $n_{1/2}$ that appears both in the skyrmion$_{d\pi}$ crystal simulation and in the closure approximation with the tensor forces. 
}\label{Esym-cusp}
\end{figure}

Consider the symmetry energy $E_{sym}$ in the energy per baryon of nuclear matter 
\be
E(n,\zeta)=E(n,0)+E_{sym}(n)\zeta^2+O(\zeta^4)+\cdots\label{Esym}
\ee
where $\zeta=(N-Z)/(N+P)$ with $N(Z)$ being the neutron (proton) number. In the skyrmion$_{d\pi}$, while one cannot calculate the energy per baryon reliably, $E_{sym}$ can be calculated in the large $N_c$ limit by collectively rotating the whole skyrmion matter. For $A=N+Z$ system with $N\gg Z$, it is found~\cite{PLR-Esym}  
\be
E_{sym}\approx (8 \lambda_I)^{-1}
\ee
where $\lambda_I$ is the isospin moment of inertia. There are two contributions, one  from the leading current algebra ($O(p^2)$) term and the other from the Skyrme quartic ($O(p^4)$) term. Both turn out to be equally important. The predicted $E_{sym}$~\cite{PLR-Esym} is  schematically given in Fig.~\ref{Esym-cusp}.

What is notable here is the cusp at what corresponds to the skyrmion-to-half-skyrmion transition density $n_{1/2}$. Approaching from below the density $n_{1/2}$  at which the cusp is located, the symmetry energy {\it decreases} and it increases after $n_{1/2}$ almost linearly in density. It is easy to understand how this cusp appears and it has a crucial impact on theEoS of dense matter. It arises from the interplay between the two terms with $\bar{\Sigma}=0$ figuring importantly.  The cusp cannot be present without the quartic term.

Now the question is: What could this mean in terms of the effective Lagrangian $\psi d$HLS? The answer lies in that the Skyrme quartic term represents heavy degrees of freedom, inherited from the lowest-lying $\rho$ meson and higher towers of isosvector vector mesons integrated out in the Lagrangian.  It suffices to limit to the lowest $\rho$ meson and ask how it can contribute to the symmetry energy. In standard EFT approaches, one can address it in terms of nuclear potentials given in the $\psi d$HLS Lagrangian. It is well-known that the symmetry energy can be fairly well approximated by the closure approximation of the two-nucleon diagrams with iterated tensor forces,
\be
E_{sym}\approx  \kappa \frac{\la (V_T)^2\ra}{\Delta E},\label{closure}
\ee
where $\kappa > 0$ is a known constant and $\Delta E\approx (200 - 300)$ MeV is the energy of the intermediate states to which the tensor force $V_T$ connects dominantly from the ground state of the matter.  It is also very well known in nuclear theory that the net tensor force gets contributions from one-pion and one-$\rho$ exchanges with opposite signs, so the dominant one-pion exchange  tensor force,  attractive in the range of two-nucleon interactions,  is weakened by the interference of the two terms. In medium the pion mass stays more or less unaffected by density but the $\rho$ mass decreases with density with the decreasing dilaton condensate (which is related to the pion decay constant)  -- \`a la Brown-Rho scaling which holds in $\psi d$HLS -- so the cancellation in the tensor force reducing the net strength would make the symmetry energy ultimately vanish at some density above $n_0$. This would be  strongly at odds with what is known in nuclear physics at $\sim n_0$ and slightly above via heavy-ion experiments. This (potential) disaster can be avoided if at some density the gauge coupling $g_\rho$ drops faster than the pion decay constant does. By imposing this condition at  $n_{1/2}$, one can make the $\rho$ tensor drastically suppressed so as to let the pion tensor force take over and make the cusp structure develop at $n_{1/2}$. This reproduces qualitatively the same cusp in Fig.~\ref{Esym-cusp}. Indeed  the HLS coupling $g_\rho$ going  to zero  so that $m_\rho\propto g_\rho\to 0$  at the VM fixed is the basic premise of the HLS scheme which is in consistency with the Suzuki theorem.  This is the first -- and the most important -- input from topology in the parameters of $\psi d$HLS.\footnote{We mention as a side remark that this scenario invalidates  the notion made in some heavy-ion circles working on dilepton processes that chiral restoration at high temperature implies degeneracy in mass between $\rho$ and $a_1$, not the mass going to zero (in chiral limiet).}  

\begin{figure}[htbp]
\begin{center}
\includegraphics[width=0.4\textwidth]{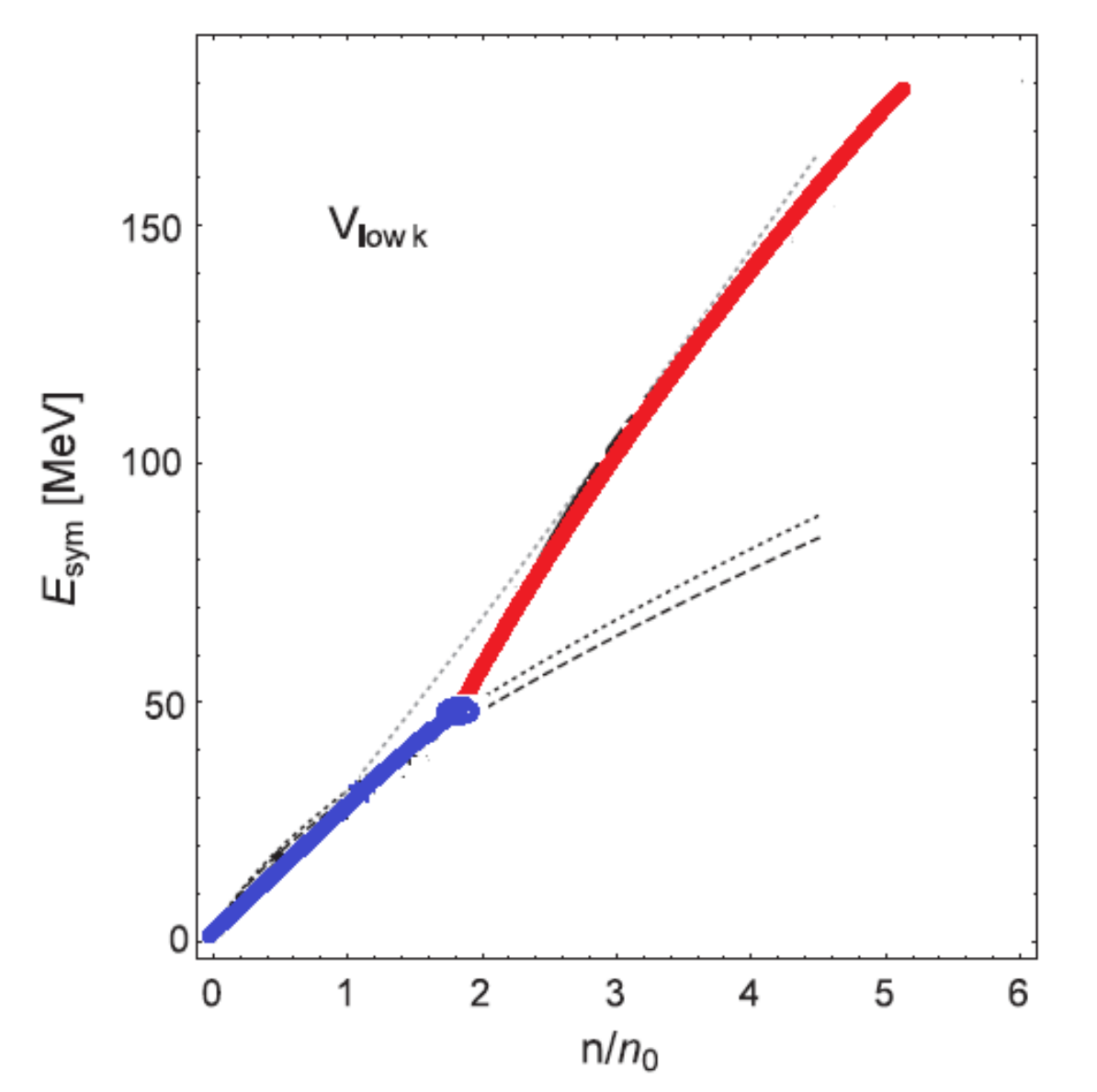}
\end{center}
\vskip -.5cm
\caption{$E_{sym}$ in $V_{lowk}$ RG in $Gn$EFT for $n_{1/2}\sim 2n_0$ copied from \cite{PKLMR}. The theoretical points given in $V_{lowk}$ RG are connected by blue line below $n_{1/2}$ and by red above $n_{1/2}$. The dotted lines indicate experimental constraints.} \label{EsymVlowk}
\end{figure}

What's given in Fig.~\ref{Esym-cusp} corresponds roughly to mean-field approximations valid near $n_{1/2}$. The cusp singularity structure must be an artifact of the approximation involved in both the skyrmion crystal and the closure approximation, involving large $N_c$ and large $\bar{N}$ limits. Such a cusp structure must be smoothed by fluctuations on top of the mean field. In \cite{PKLMR}, corrections to the mean field treated in terms of $1/\bar{N}$ corrections \`a la $V_{lowk}$RG do indeed smooth the ``singularity" but significantly leave unmodified the soft-to-hard cross-over at $n_{1.2}$. This is shown in Fig.~\ref{EsymVlowk}.

We will return to this cusp structure in Sec.~\ref{tension} regarding the structure of $E_{sym}$ inferred from the PREX-II experiment in $^{208}$Pb~\cite{PREX-II}.

\subsection{Soft-to-hard crossover in EoS}\label{soft-to-hard}
The cusp structure in $E_{sym}$ discussed above have a crucial impact on the EoS of massive compact stars. It describes {\it softening} of the EoS going toward $n_{1/2}$ from below and {\it  hardening} going above $n_{1/2}$. This feature, which will be elaborated in what follows, is consistent with what is observed in the tidal polarizability (TP) $\Lambda$ from gravity waves and also with the massive $\gsim 2 M_\odot$ stars. The former requires a soft EoS below $n_{1/2}$ and the latter a hard EoS above $n_{1/2}$, both of which are naturally offered by the cusp structure. 
{\it This cross-over behavior is the main reasoning for identifying the skyrmion-half-skyrmion transition with hadron-quark continuity.}
\subsection{Scale-invariant quasiparticles}\label{quasiparticles}
Another striking feature in the skyrmion-half-skyrmion transition is that at $n_{1/2}$ the confined half-skyrmion complex behaves like non-interacting fermion, i.e., ``free quasiparticles."

\begin{figure}[h]
\begin{center}
\includegraphics[width=4.4cm]{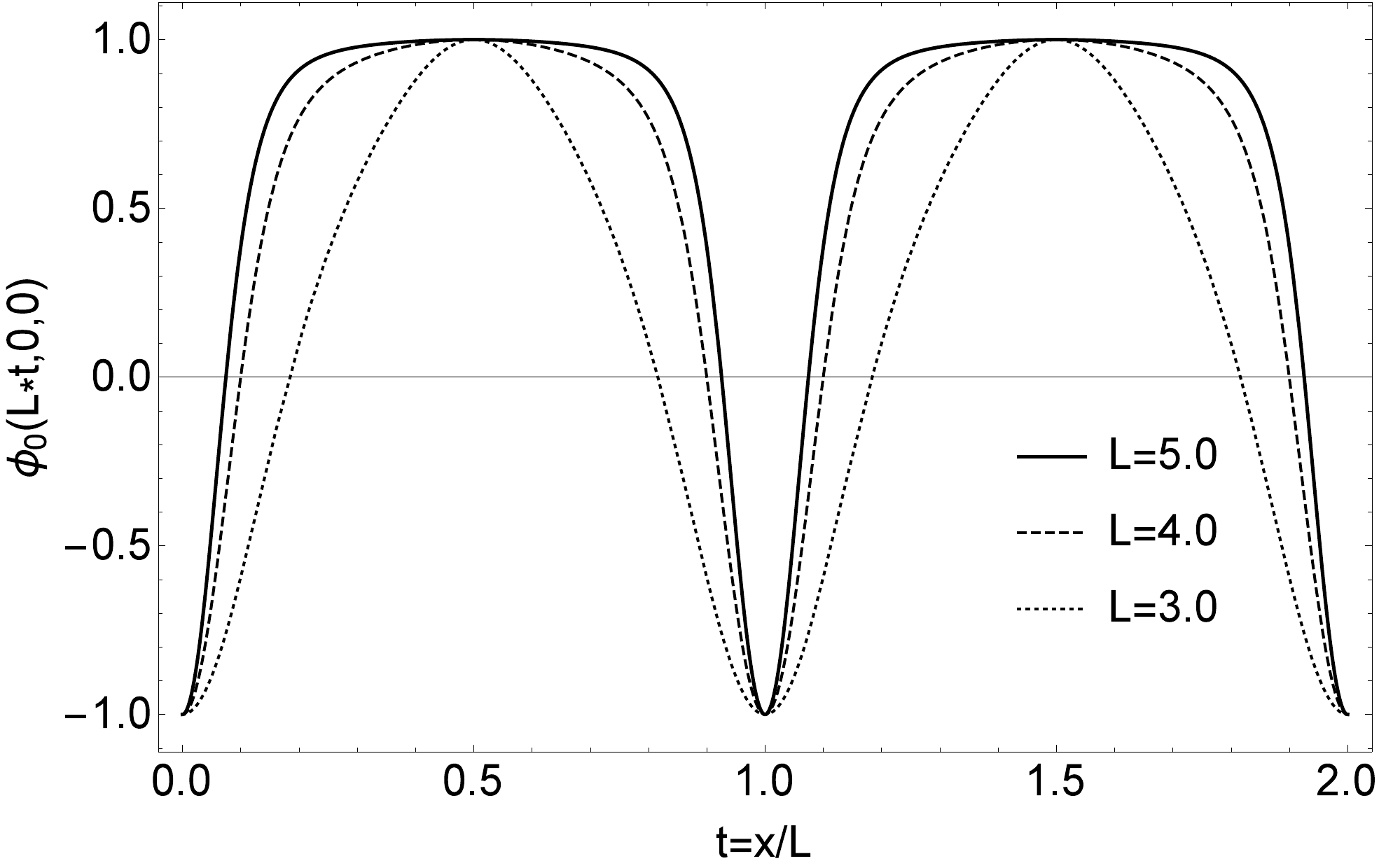}\includegraphics[width=4.4cm]{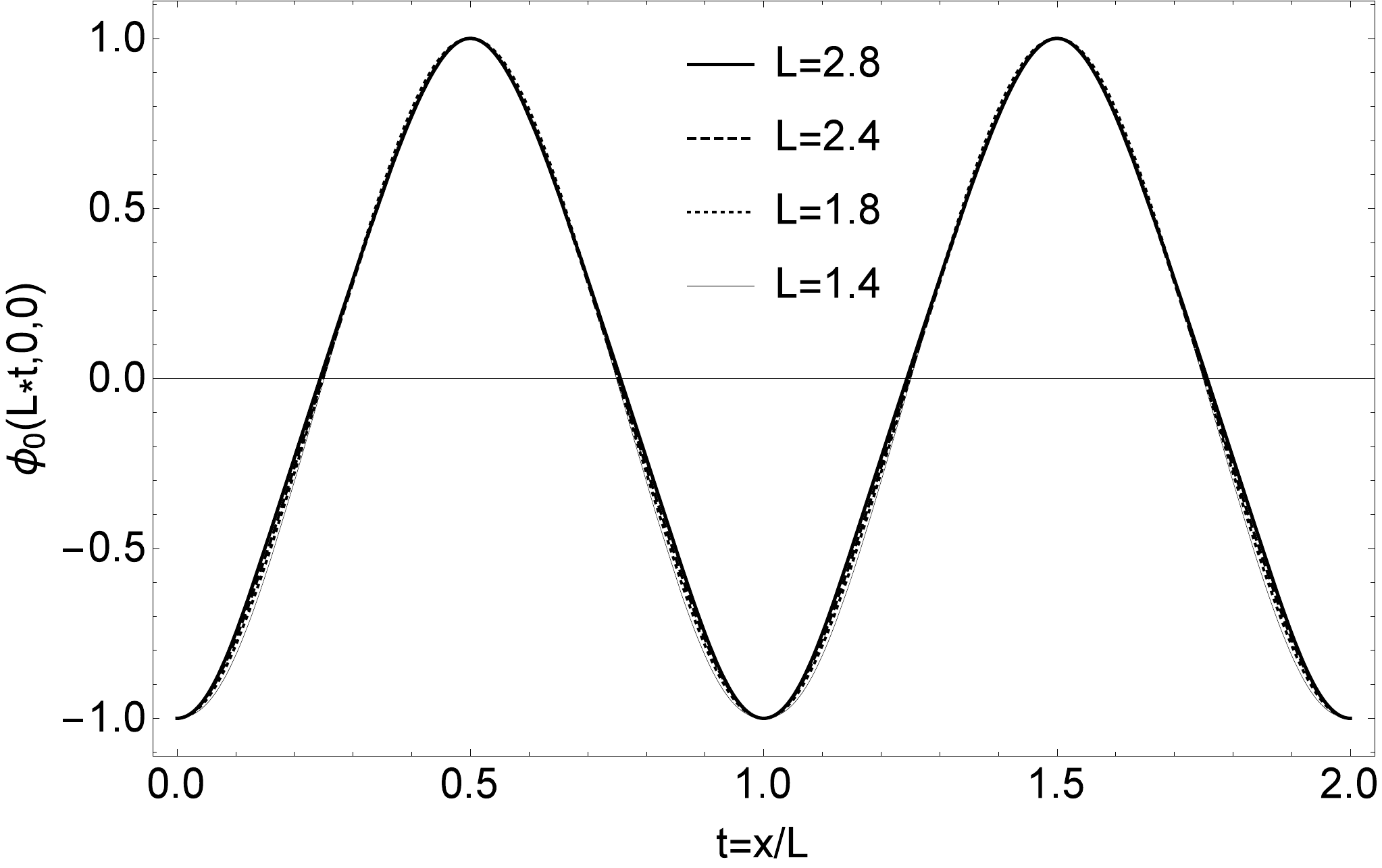}
\includegraphics[width=4.4cm]{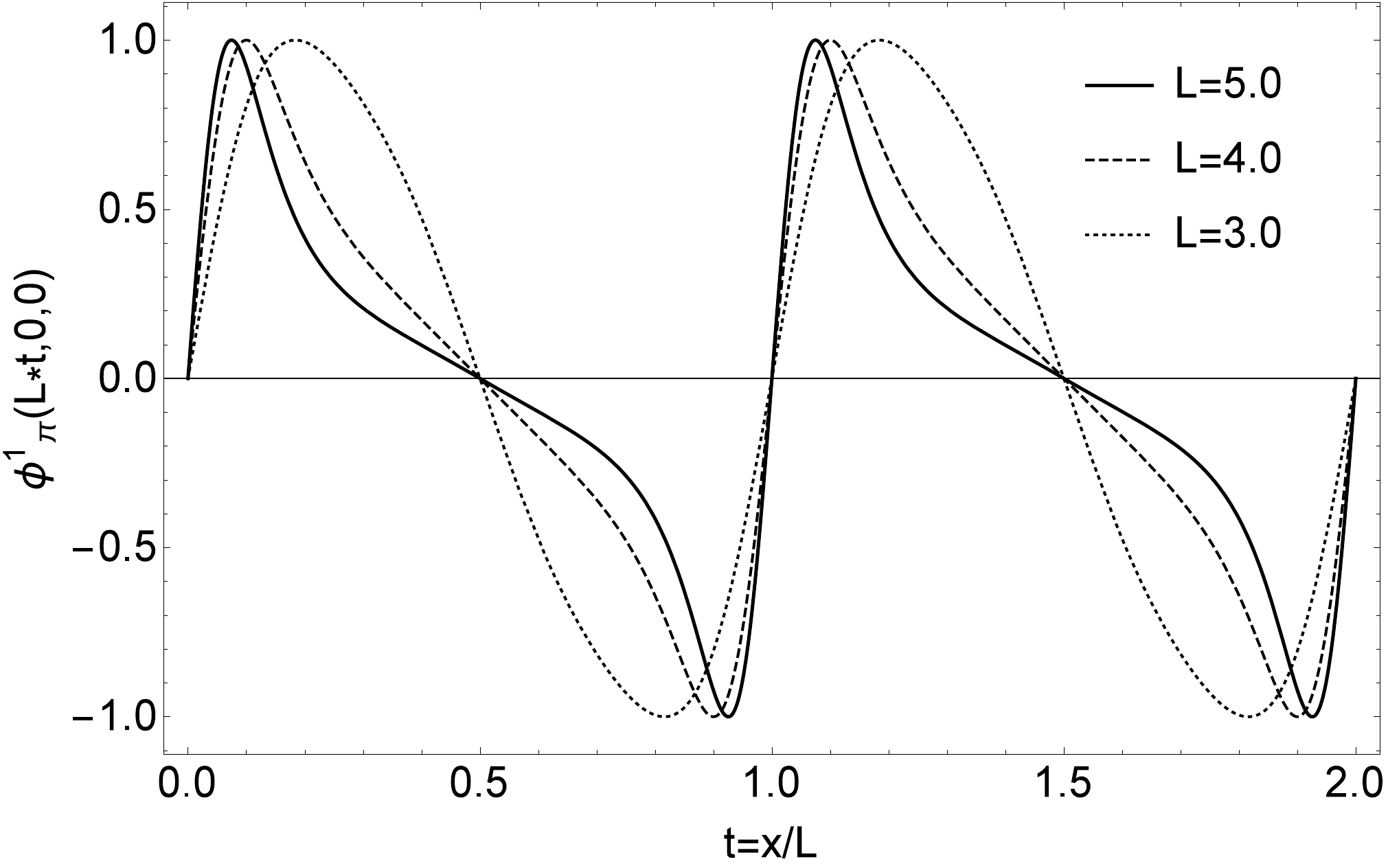}\includegraphics[width=4.4cm]{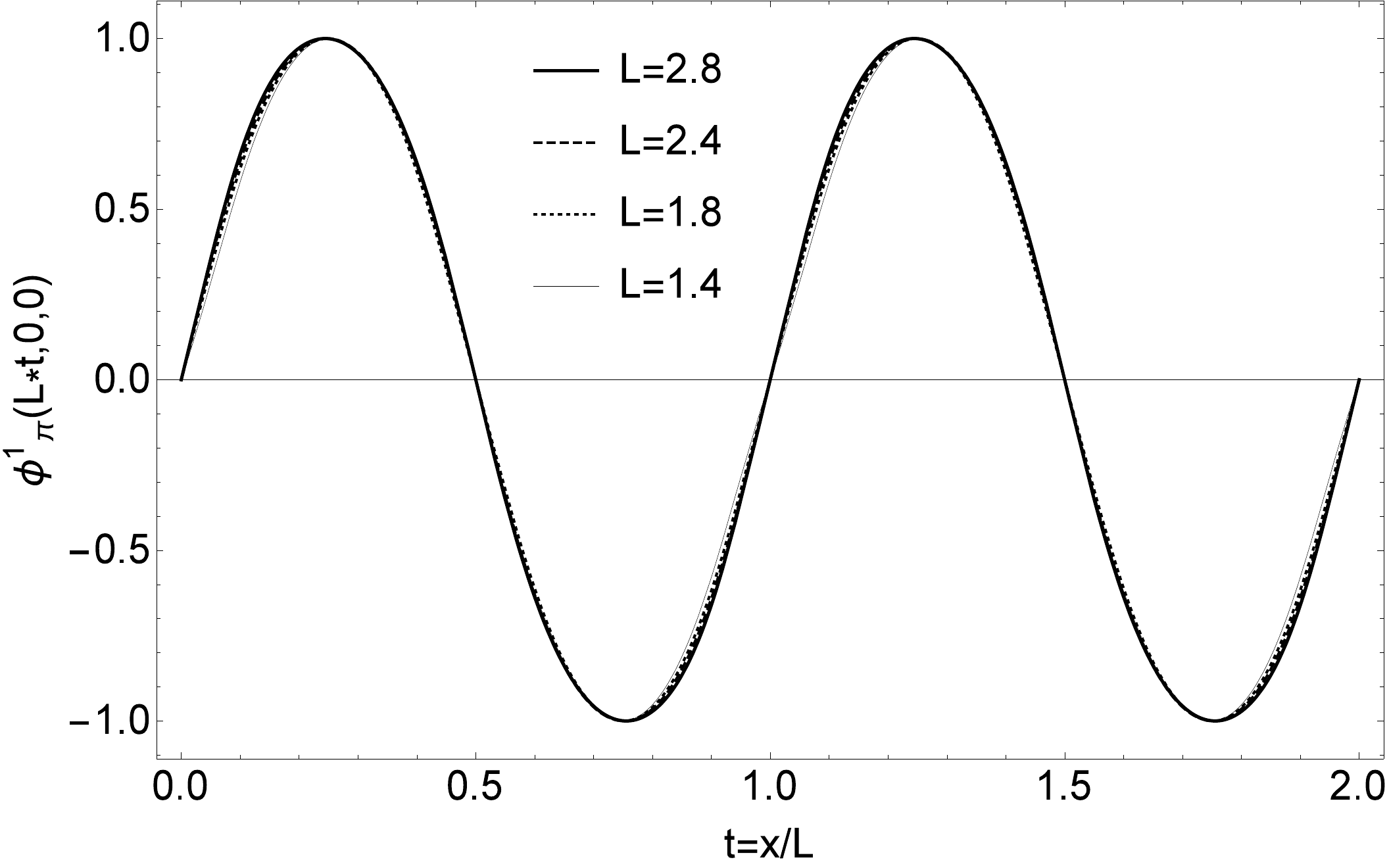}
\caption{  The field configurations $\phi_0$ and $\phi^1_\pi$  as a function of $t = x/L$ along the y = z = 0 line. The maximum values for $\eta=0,\pi$ are $\phi_{0,\,L,\, {\rm max}} = \phi_{\pi,\,L,\, {\rm max}} = 1$. The left and right panels correspond respectively to the skyrmion phase and half-skrymion phase. The half-skyrmion phase sets in when $L=L_{1/2} \lsim 2.9\,{\rm fm}$ which should not be identified with physical velue.}
\label{quasi}
\end{center}
\end{figure}

Let us write the chiral field as
\be
U(\vec{x})=\phi_0 (x,y,z) +i\phi_\pi^j (x,y,z)\tau^j
\ee
and the fields placed in the lattice size $L$ as $\phi_{\eta,L} (\vec{x})$ for $\eta=0,\pi$ and normalize them with respect to their maximum values denoted $\phi_{\eta,L,max}$ for given $L$. The properties of these fields are plotted for $L>L_{1/2}$ -- skyrmion phase -- and $L\leq L_{1/2}$ -- half-skyrmion phase -- in Fig.~\ref{quasi}~\cite{MR-review}.
Note the remarkable feature that the field configurations in the half-skyrmion phase are density-independent in strong contrast with the skyrmion phase. This transition is interpreted to signal the onset of pseudo-conformal invariance in baryonic matter to be observed in the pseudo-conformal sound velocity in massive compact stars discussed in Sec.~\ref{compact-stars}.
\section{$Gn$EFT}
We are now adequately equipped to specify the structure of $Gn$EFT built from $\psi d$HLS Lagrangian. 

We will work with  $\psi d$HLS defined in the ``vacuum" sliding with density. The parameters of the Lagrangian will therefore depend on density. With the input taking into account the topology change at $n_{1/2}$, the parameters will change qualitatively as density goes from below to above $n_{1/2}$. 

There are several ways by which the dependence can occur.

First the conformal compensator field will pick up the VEV, $\chi\to \la\chi\ra_n +\chi$ as density changes, so the parameters of the Lagrangian term coupled to $\chi$ will depend on density via $\la\chi\ra_n$. 

Next  the matching of correlators between $\psi d$HLS and QCD~\cite{HY:PR}  at a given density will introduce density dependence, such as the NG boson decay constants and coupling constants $g_V$, $g_{\chi NN}$ etc., a most important quantity of which being the $\rho$ coupling $g_\rho$ in medium which flows to  $g_\rho=0$ as density goes to the VM fixed point. 

The density dependence on parameters will be denoted by superscript $\ast$, e.g., $m^\ast$ for masses, $g^\ast$ for coupling constants, $\la (\chi,\bar{q}q) \ra^\ast$ for comdensates  etc.

Now given the $\psi d$HLS endowed with scaling parameters encoding the topology change at $n_{1/2}$, we adopt to approach nuclear dynamics in RG-implemented Fermi-liquid as announced in Introduction\footnote{We could in principle formulate chiral-scale-HLS perturbation theory along the line of nuclear $\chi$PT. This has not been formulated yet.}  We will discuss first Landau Fermi-liquid (FL) fixed point approach and, whenever available, the $V_{lowk}$RG approach that goes beyond the FL fixed point. The details of the calculations given in \cite{MR-review} will be skipped. We go immediately into confronting Nature.
\subsection{Dilaton-limit fixed point}
The first application of $Gn$EFT is to go to high density in the mean-field approximation with $\psi d$HLS.

As noted, the mean-field approximation with $\psi d$HLS is equivalent to Fermi-liquid fixed point approach valid in the large $N_c$ and large $\bar{N}$ approximation~\cite{BR91,FR}. As suggested in Sect~\ref{quasiparticles}, this approximation should be more appropriate at high density. For simplicity we take the LOSS approximation for scale symmetry described in Sec.~\ref{LOSSA}.

Following Beane and van Kolck~\cite{bira}, we do the field re-parametrization  ${\cal Z}=U\chi f_\pi/f_\chi=s+i\vec{\tau}\cdot\vec{\pi}$ in $\psi d$HLS in LOSS and take the limit ${\rm Tr} ({\cal Z}{\cal Z}^\dagger)\to 0$. Two qualitatively different terms appear from the manipulation:  one regular and the other singular in the limit. The singular part is of the form
\be
{\cal L}_{\rm sing} &=& (1- g_A){\cal A} (1/{\rm Tr} \big({\cal Z}{\cal Z}^\dagger)\big)\nonumber\\ 
&+& (f_\pi^2/f_\chi^2 -1) {\cal B} \big(1/{\rm Tr} ({\cal Z}{\cal Z}^\dagger)\big).
\ee
The first (second) term is with (without) the nucleons involved. The requirement that there be no singularity leads to the dilaton-limit fixed point (DLFP) ``constraints"
\be
g_A\to g_A^{DL}=1\label{gAto1}
\ee 
and 
\be 
f_\pi\to f_\chi \neq 0.\label{fpifchi}
\ee
We have denoted the $g_A$ arrived at in the dilaton-limit fixed point as $g_A^{\rm DL}$ to be distinguished from the $g_A^L$ -- Eq.~(\ref{gA=1}) --  arrived at the Landau Fermi-liquid fixed point.

We note that these ``constraints" are more or less (if not exactly) the same as what are in the genuine dilaton properties approaching the IR fixed point. There are more constraints associated with  the DLFP, among which highly relevant to the EoS at high density is the ``emergence" of parity doubling in the nucleon structure (explained in detail in \cite{MR-review}). We will come to this point later as the dilaton condensate in the half-skyrmion phase $\la\chi\ra^\ast$ converges to the chiral-invariant mass $m_0$ in the parity-doubling model.  It is not known whether and how this mass  vanishes ultimately. But in our approach it can be present near the IR fixed point and more likely in the density regime relevant to compact stars. 

In what follows we will simply assume  that {\it  the DLFP is equivalent to the IR fixed point of \GDS.}

If we make the assumption that soft theorems are also applicable near the DLFP, which seems justified, then  we can write, following the discussion in Sec.~\ref{softtheorems},
\be
m_N^\ast\approx f_\chi^\ast g_{\chi NN}^\ast \approx f_\pi^\ast g_{\pi NN}^\ast/g_A^\ast
\ee
from which we have
\be
g_A^\ast\approx\Big(\frac{f_\pi^\ast}{f_\chi^\ast}\Big)\Big(\frac{g_{\pi NN}^\ast}{g_{\chi NN}^\ast}\Big).
\ee
This relation is to hold for both low and high density. At low density this relation could be verified in chiral-scale perturbation theory.  At high density $n> n_{1/2}$,  
we expect from (\ref{gAto1}) and (\ref{fpifchi}) that $g_{\pi NN}^\ast\to g_{\chi NN}^\ast$, what could be considered as a prediction of this theory.  
 \subsection{$Gn$EFT in finite nuclei and nuclear matter}
As discussed in great detail in \cite{MR-review}, the parameters of $\psi d$HLS Lagrangian for $Gn$EFT,  primarily controlled by the topology change encoding putative hadron-quark continuity,  are different from below to above $n_{1/2}$. Phenomenology fixes the range $2\lsim n_{1/2}/n_0 \lsim 4$. As a typical value, we will use $n_{1/2}\approx 3n_0$. The recent development suggests $n_{1/2}$ could be higher as we will comment later.

Extended to a higher cutoff scale with heavier degrees of freedom than sChEFT, $Gn$EFT should, by construction, go beyond the scale applicable to sChEFT for baryonic matter properties valid at $\sim n_0$~\cite{HRW}.  With the heavy vector and scalar degrees of freedom explicitly figuring in the dynamics, the power counting rule, however, is considerably different from that of sChEFT in which the vector and scalar excitations enter at loop orders. Instead of doing the standard EFT based on power counting involving the vector mesons and the dilaton, we opt to approach the problem in Wilsonian RG-type strategy along the line worked out for strongly correlated condensed Fermi-liquid systems~\cite{shankar}.  At the mean-field level, this procedure gives the Fermi-liquid fixed point theory~\cite{FR} and going beyond the mean field in $V_{lowk}$RG, it has been verified that {\it all} thermodynamics properties of nuclear matter come out as well as in sChEFT treated up to N$^4$LO--  but with much fewer number of fit parameters~\cite{MR-review}. That the extremely simple $Gn$EFT calculation works as well as the high chiral-order sChEFT calculation may be reflecting the ``magic" of HLS (Seiberg-)dual to the gluon of QCD as recently noted by several authors~\cite{Komargodski,karasik,karasik2}.

We stress here that the $Gn$EFT applied  to nuclear matter properties at $\sim n_0$ is in the LOSS approximation described in Sec.~\ref{LOSSA}. It has not been yet checked how the leading order (LO) chiral-scale Lagrangian without the LOSS approximation -- with the possible role of  $\beta^\prime$ dependence -- affects the results . This point is pertinent in connection with the quenched $g_A$ problem to be treated next where hidden scale symmetry gets ``un-hidden" together with a possible influence of the anomalous dimension $\beta^\prime$. Also note that the result that the standard high-order sChEFT and the $Gn$EFT with explicit hidden symmetries fare equally well at $n\sim n_0$ indicate that those symmetries are buried and not apparent in the EOS at that density. This does not mean all observables in nuclear processes at low density are opaque to them. Indeed totally unrecognized in the past  is that the long-standing mystery of quenched $g_A$ in light nuclei as seen in shell models  has a deep connection with how scale symmetry hidden in QCD emerges in nuclear correlations~\cite{gA}. In fact this quantity exhibits the multifarious ways hidden scale symmetry manifests in these low-energy nuclear processes. 
\subsubsection{Quenching of $g_A$ in nuclei}
It has been a long-standing mystery that simple shell-model calculations of the superallowed Gamow-Teller (GT) transition in light nuclei required  an effective GT coupling constant $g_A^\ast\approx 1$, quenched from the free-space value 1.267, to explain the observed decay rates~\cite{gA-shellmodel}. There have been a variety of explanations of the mystery debated in the literature since 1970's but none with convincing arguments. Here we offer a mechanism which seems to resolve the half-a-century mystery: It involves an up-to-date unrecognized working of hidden scale symmetry in nuclear interactions~\cite{gA}.   This mechanism has a remarkable impact  not only on the structure of dense compressed baryonic matter in compact stars but also on how scale (or conformal symmetry) is realized in nuclear medium.

We first give precise definitions for addressing nuclear Gamow-Teller transitions both with and without neutrinos.  What has been referred to in the literature as ``quenched $g_A$" in (supperallowed) GT transitions in light nuclei treated in simple shell models~\cite{gA-shellmodel} is a misnomer. Most, if not all, of the so-called ``quenching" involved in bringing the $g_A$ effective in nuclei -- denoted generically as $g_A^\ast$ --  have little to do with a ``genuine quenching" of the axial-vector coupling constant in the weak current. In fact $g_A^\ast\approx 1$ will be shown to signal the emergence of scale symmetry hidden in nuclear interactions. It seems to be connected to the dilaton-limit fixed point $g_A^{\rm DL}=1$ at high density discussed above, Eq.~(\ref{gAto1}).

 In order to clear up the above confusion and zero-in on the issue concerned, we write  
the leading chiral-scale-order axial current $A_\mu ^{\pm}$ in $\psi d$HLS  in medium as 
\be
A_\mu^{\pm} = g_A q_{\rm ssb} \bar{\psi} \tau^\pm \gamma_\mu\gamma_5\psi +\cdots\label{axialcurrent}
\ee 
where
\be
q_{\rm ssb}=c_A+(1-c_A){\Phi^\ast}^{\beta^\prime}, \ \Phi^\ast=f_\chi^\ast/f_\chi\label{qssb}
\ee
where $q_{\rm ssb}$ represents the quantum-anomaly-induced scale-symmetry-breaking  (ssb). Note that in the \GDS, $q_{\rm ssb}$ when $\neq 1$  is a generically density-dependent multiplicative factor on the scale-invariant axial current $g_A  \bar{\psi} \tau^\pm \gamma_\mu\gamma_5\psi$. As mentioned in Sec.~\ref{CSS}, $c_A$ is  an unknown constant. The LOSS approximation corresponds to taking $q_{\rm ssb}\approx 1$. For what follows, it is important to note that what amounts to the true coupling-constant quenching of  $g_A$ -- in the sense of renormalization inherited from QCD -- is 
\be
\bar{g}_A\equiv g_A d_{\rm ssb}.
\ee 
This is the quantity that should be identified as the bona-fide quenched $g_A$ in nuclear medium and that we aim to extract from GT transitions in nuclei.

First we will calculate $g_A^\ast$ for the LOSS with $d_{\rm ssb}=1$.  We would like to see how the scale invariance of the axial current impacts on the nuclear GT matrix element where nuclear correlations are treated in the LOSS approximation.

In the large $N_c$ and large $\bar{N}$ limit  an astute way to approach is  to apply the Goldberger-Treiman relation in nuclear medium as suggested in Sec.~\ref{softtheorems}. This, as worked out in \cite{FR} for the axial current (\ref{axialcurrent}) with $q_{\rm ssb}=1$, is of the form~\cite{BR91,CND}\footnote{As is well-known, calculating the $g_{\pi NN}$ coupling in the skyrmion model is highly involved, so what was done in Adkins, Nappi and Witten~\cite{ANW} was to use the Goldberger-Treiman relation to arrive at that coupling. We are using the same strategy to the in-medium Goldberger-Treiman relation exploiting the scale invariance of the Skyrme quartic term.  (See  \cite{CND}).}
\be
(g_A^{\rm eff}/g_A)^{d_{ssb}=1}\equiv g_A^L/g_A&=& (\Phi^\ast)^{-2} (m_N^L/m_N)^2 \nonumber\\
&+& O(1/N_c, 1/\bar{N})\label{FLpred}
\ee
where the superscript $L$ represents density-dependent Fermi-liquid fixed point quantities. $g_A^L$ depends on density via $\Phi^\ast$ and $m_N^L$. We note  that  $g_A^{\rm eff}$  is the ``effective" axial-vector coupling constant for the {\it quasiparticle} treated  in Fermi-liquid approach on the Fermi surface,  which is not to be associated with the $g_A^\ast$ as has been done in  the literature for a {\it true quenched coupling constant} applicable generically to {\it all} axial transitions . To what it corresponds will be precisely defined below.  

In the RG approach, the Landau mass is {\it taken} as a fixed point quantity~\cite{shankar}. That the axial coupling constant is also taken as a fixed point quantity is something new and unfamiliar in the field. What it implies will be explained.

There are other quantities in nuclear electroweak responses where the same arguments based on soft theorems in the large $N_c$and large $\bar{N}$ work very well. 

We now explain what $g_A^L$ signifies and how it is connected to what is computed in shell-model calculations. 

We are concerned only with the superallowed Gamow-Teller (GT) transition which involves zero momentum transfer. We assume that the matrix element is not accidentally suppressed. Now in the Wilsonian-type RG equation for strongly correlated fermion (nucleon here) systems, decimated all the way to the top of the Fermi sea, the GT transition from, say, a quasi-proton to a quasi-neutron on top of the Fermi sea is given  by the GT matrix element 
\be
{\cal M}_{GT}^{d_{ssb}=1}= g_A q^L (\tau^- \mathbf{\sigma})_{fi} \label{FLM}
\ee
where all nuclear correlations are captured entirely in what we denote as $q^L$ standing for Landau Fermi-liquid fixed point.  

Before discussing the prediction of this approach, let us explain in what way this matrix element is related to shell-model calculations.  To be specific, we consider the doubly magic nucleus $^{100}$Sn with 50 neutrons and 50 protons. We focus on this nucleus because it  has  what may be up-to-date the most accurate data on the superallowed GT transition. This process allows to exploit the ``extreme single-particle shell model (ESPSM)" description~\cite{ESPSM,RIKEN} which provides a precise mapping of the shell-model result to the Fermi-liquid fixed point result (\ref{FLM}). In the ESPSM description, the GT process involves the decay of a proton  in a completely filled shell $g_{9/2}$ to a neutron in an empty shell $g_{7/2}$.  We write the GT matrix element in the ESPSM description in the form
\be
{\cal M}_{\rm GT}^{espsm}= g_A q^{\rm ESPSM} (\tau^- \mathbf{\sigma})_{fi}\label{ESPSM}
\ee
where $q^{\rm ESPSM}$ is the ratio of the {\it full} GT matrix element  given in principle by nature (that is to be hopefully captured in an accurate experiment) to the accurately calculable ESPSM matrix element for the GT transition of the axial-current with $q_{\rm ssb}=1$. We are taking the RIKEN experiment~\cite{RIKEN} to provide the presently available empirical value of $q^{\rm ESPSM}$. Our proposal is that the Landau fixed-point description is equivalent to the ESPSM description
\be
 q^L=q^{\rm ESPSM}/q_{\rm ssb}.\label{qL=qESPSM}\label{equivalence}
\ee

Now to access $q^L$, we first note that the Fermi-liquid fixed point prediction (\ref{FLpred}) requires the Landau mass for the nucleon. Unlike in the Landau Fermi-liquid theory for electrons in which the fixed-point two-body interactions are all local,  the presence of Nambu-Goldstone bosons requires a nontrivial input, The  pion exchange contribution to the Landau(-Migdal) interaction (for nuclear physics) brings in a non-local term. This term is of $O(1/\bar{N})$,  but it is extremely important  due to the special role of soft theorems in low-energy nuclear dynamics as stressed above. The result for $q^L$,  obtained a long time ago~\cite{BR91,FR}, turns out to be  surprisingly -- and deceptively -- simple,
\be
q^{L}=\Big(1-\frac 13 \Phi^\ast\tilde{F}_1^\pi\Big)^{-2}.\label{fixedFL}
\ee
In $Gn$EFT the dilaton decay constant $f_\chi^\ast$ is locked to the pion decay constant $f_\pi^\ast$. The latter has been measured in deeply bound pionic nuclear experiments~\cite{yamazaki}, so is known as function of density.  The pionic Landau parameter $\tilde{F}_1^\pi$ is given by the Fock diagram, so is calculable for given densities. Thus both $\Phi^\ast$ and $\tilde{F}_1^\pi$ are accurately known for any density in the vicinity of $n_0$.  Furthermore the product $ \Phi^\ast\tilde{F}_1^\pi$ turns out to be surprisingly independent of density, so (\ref{fixedFL}) applies to light as well as heavy nuclei. At $n\approx n_0$, we have 
\be
q^L\approx 0.79 - 0.80.\label{qL}
\ee
This gives 
\be
g_A^L=g_A q^L\approx 1.0.\label{gA=1}
\ee
There are caveats in this result, among which the numerical value of (\ref{qL}) depends on the accuracy of the Landau Fermi-liquid fixed point approximation (the large $N_c$ and large ${\bar{N}}$ limit),  the ``precise" value of density, the  coupling constants used etc., so it would be unwise to take (\ref{gA=1}) too literally.  Ultimately this formula could be checked by a high-order chiral-scale expansion as is done with some soft theorems in nuclear matter discussed in Sec.~\ref{softtheorems}. One can however have  confidence in this result in that the same soft theorems applied to nuclear EM response functions at low energy appear to work extremely well. An example is the anomalous orbital gyromagnetic ratio of the proton $\delta g_L^p$~\cite{FR} 
\be
\delta g_l=\frac{4}{9} [(1/\Phi^\ast)-1-\frac 12 \tilde{F}_1^\pi] \tau_3.
\ee
This involves the same quantities as in $g_A^L$, $\Phi^\ast$ and $\tilde{F}_1^\pi$ and  follows from the same soft theorems in the LOSS approximation as in (\ref{gA=1}).  The prediction for the Pb nucleus $\delta g_L^p= 0.22\tau_3$ is in good agreement with the presently available experiment $(\delta g_l^p)_{exp} =  0.23\pm 0.03$~\cite{FR}.\footnote{This should be contrasted with the presently available result in sCHEFT~\cite{HRW} which disagrees with the empirical value by more than factor of 3.} We take this as an indication of the reliability of (\ref{fixedFL}).

Now what does $g_A^L\approx 1$ mean?

What this means is that thanks to the equivalence (\ref{equivalence}) an accurate Gamow-Teller matrix element calculated in a full-scale shell-model with the axial current operator (\ref{axialcurrent}) with $q_{\rm ssb}=1$  should correspond to the FLFP matrix element. Of course at the present status of computational power, such a full scale shell-model calculation is not doable for the double-closed-shell nucleus  $^{100}$Sn. However in light nuclei, one expects $q_{\rm ssb}\approx 1$  because $\Phi^\ast$ as well as the $c$ coefficients could be near the vacuum values. What Eq.~(\ref{qL=qESPSM}) implies is that high-quality calculations such as  quantum Monte Carlo method should reliably explain the Gamow-Teller transitions {\it without} quenching of $g_A$ from the free-space value.  We suggest that this expectation is supported by the recent (powerful) quantum Monte Carlo calculations for $\beta$ decay and electron capture in $A= 3-10$~\cite{wiringa}  that  can explain the measured experimental values at $\sim 2\%$ uncertainty with the unquenched $g_A=1.276$ and without multi-body exchange currents. 

It is worth noting that the result (\ref{gA=1}), predicted already in early 1990, takes a totally new aspect to it.  

At the risk of being redundant, let us recall that Eq.~(\ref{gA=1}),  justified in the large $N_c$ and large $\bar{N}$ limits,  corresponds to the Landau FLFP that exploits soft-dilaton theorems~\cite{GDS}.   As formulated it can be precisely equated to the superallowed  Gamow-Teller transition matrix element given in the extreme single-particle shell-model  doubly magic nuclei, e.g.,   $^{100}$Sn. At the matching scale to QCD,  the axial weak coupling to the nucleons is scale-invariant, hence the renormalization leading to (\ref{gA=1}) can be considered {\it entirely due to nuclear (many-body) interactions taking place below the matching scale}. Thus (\ref{gA=1}) is capturing the influence of the putative scale symmetry  emergent from the (nuclear) interactions  which may (or may not) be intrinsically connected to QCD. We can see this  also by going to high density $\gg n_0$. Starting with non-linear sigma model with constituent quarks, as shown by  \cite{bira},  letting the dilaton mass $m_\chi$ go to zero with the conformal anomaly turned off --- and in the chiral limit --- leads to a linearized Lagrangian that satisfies various well-established sum rules based on soft theorems,  provided the singularities that appear as $m_\chi\to 0$ are suppressed by the dilaton-limit  fixed-point  constraints. There is a ``continuity" from low density to high density in the way scale symmetry (combined with hidden local symmetry) manifests itself in strong interactions.
\subsubsection{Evidence for $d_{ssb}\neq 1$?}
Up to density in the vicinity of $n_0$, there is no indication for corrections to the LOSS approximation $q_{\rm ssb}\approx 1$. The overall properties of nuclear matter are consistent with the LOSS approximation. But there is no strong argument that the $c$ coefficients be $\approx 1$ or $\beta^\prime\ll 1$ with $\Phi^\ast\neq 1$. In fact treating dense matter in terms of skyrmions on crystal lattice is found to give a sensible result~\cite{omega} only if $\beta^\prime\sim (2-3)$ and $c_{\rm hWZ}\approx 0$ where $c_{\rm hWZ}$ is the $c$ coefficient in the $q_{\rm ssb}$ accompanying the homogeneous Wess-Zumino term in the HLS Lagrangian~\cite{HY:PR}. Otherwise the system becomes catastrophically unstable unless the $\omega$ meson mass goes to $\infty$. This would of course be a nonsense. It could just be an indication for anomalous behavior in the high density regime which is not understood at all,  but it seems totally unnatural.

Given what appears to be an accurate result for the superallowed Gamow-Teller transition for the doubly magic nucleus $^{100}$Sn, one can raise the question as to what one can learn about the correction  to the LOSS, namely, the deviation from $q_{\rm ssb}=1$.  Now to extract $q_{\rm ssb}$ from  the RIKEN experiment, we write the $^{100}$Sn matrix element expressed in terms of the ESPSM
\be
{\cal M}_{\rm GT:riken}= g_A q_{\rm riken}^{\rm ESPSM} (\tau^- \mathbf{\sigma})_{fi}.\label{RIKEN} . 
\ee
The RIKEN experiment leads to 
\be
q_{\rm riken}^{\rm ESPSM}= 0.46 - 0.55.\label{rikenquenching}
\ee
From the equivalence relation (\ref{equivalence}), one can get
\be
q_{\rm ssb}^{\rm RIKEN}\approx 0.58-0.69.\label{qssbRIKEN}
\ee
The difference $(1-q_{\rm ssb}^{\rm RIKEN})$ represents the anomaly-induced correction to the LOSS. This is substantial. 

Next given the experimental (RIKEN)  information on $q_{\rm ssb}$, (\ref{qssbRIKEN}), one might hope to get an idea on $\beta^\prime$. Since $c_A$ is not known, one cannot zero-in on a unique answer. Let us however see what one gets if one just takes $\beta^\prime\approx 2.5$ and $c_A\approx 0.15$ compatible with what's indicated for the hWZ term in the skyrmion crystal model~\cite{omega}. That gives $q_{\rm ssb}\approx 0.62$ which is in the ball-park of the RIKEN value for $q_{ssb}$, (\ref{qssbRIKEN}).  

There is a caveat here. The skyrmion crystal simulation involves high density for which $\Phi^\ast \ll 1$ whereas the RIKEN process involves $\Phi^\ast\lsim 1$. There is no reason why the same value of the $c$ coefficient makes sense.

It would therefor be desirable to measure $q^{\rm ESPSM}_X$ for different double-magic systems $X$ in addition to reconfirming the RIKEN data. Certain forbidden axial transitions treated in the next subsection offer a promising possibility.  
\subsubsection{Forbidden axial transitions}
The deviation from the LOSS approximation encoding the scale symmetry breaking (\ref{qssb}), if firmly confirmed, implies that the $g_A^\ast\approx 1$ observed in simple shell model calculation in light nuclei is not to be identified with an effective $g_A$ applicable to {\it all} axial-vector processes in nuclear medium. For example, it cannot be applied to such processes as neutrinoless double $\beta$ decays -- relevant for going BSM -- where momentum transfers can be of $\sim 100$ MeV.  It is certainly not a ``fundamentally renormalized" coupling constant generically applicable to {\it all} processes in nuclei.  This issue is recently raised in the analysis of effective $g_A$,  denoted $\bar{g}_A$,  in the $\beta$ decay  spectrum-shape function involving leptonic phase-space factors and nuclear matrix elements in forbidden non-unique beta decays~\cite{cobra} -- referred to as COBRA in what follows.  The nuclear operators involved there are non-relativistic momentum-dependent impulse approximation terms. In principle there can be  $n$-body exchange-current corrections with $n \geq 2$. However the corrections to the single-particle (impulse) approximation are typically of $m$-th order with $m \gsim  3$ relative to the leading one-body term in the power counting in $\chi$EFT and could be -- justifiably --   ignored.  Unlike in the superallowed GT transition, neither soft theorems nor the Fermi-liquid renormalization-group strategy can be exploited in this work. Hence (\ref{gA=1}) is not relevant to the spectrum-shape function discussed in \cite{cobra}. In defining $\bar{g}_A$, nuclear correlations are taken into account, so what is  relevant is the possible the scale-anomaly correction (\ref{qssb}) in the analysis of \cite{cobra}. 

The  $\bar{g}_A$ gotten from the RIKEN data corresponding to the ``genuinely quenched $g_A$" can be written as
\be
\bar{g}_A^{\rm RIKEN}=g_A q_{\rm ssb}^{\rm RIKEN} \approx (0.74 - 0.88).\label{rikenbarga}
\ee
This is to be compared with the analysis of the spectrum-shape factor of $^{113}$Cd $\beta$ decay~\cite{cobra}, listed in the increasing  average $\chi$-square for three nuclear models used to calculate the forbidden non-unique decay process~\cite{cobra}
\be
&&\bar{g}_A^{\rm COBRA}=g_A q_{\rm spectshape}^{\rm COBRA}\nonumber\\
&=&0.809\pm 0.122,\ 0.893\pm 0.054,\ 0.968\pm 0.056.
\label{cobraexp}
\ee
Given the range of uncertainties involved both in the experimental data and in the theoretical models used in \cite{cobra}, what can be identified as  the anomaly-induced quenching in the COBRA spectrum-shape result (\ref{cobraexp}) is more or less consistent with the same effect  in the RIKEN's result (\ref{rikenbarga}). However $\bar{g}_A\gsim 1$ would be in tension with (\ref{qssb}) unless $\Phi^\ast$ is very close to $\Phi^\ast(n_0)$ and $\beta^\prime < 2$.

In sum, what seems to be noteworthy is that nuclear $\beta$ decay processes give an indication on scale symmetry and the anomalous dimension $\beta^\prime$ reflecting on the working of scale symmetry in strong interactions. At present there is no trustful information on $\beta^\prime$ for $N_f\sim 3$ involved in nuclear dynamics. An interesting question that cannot be answered at present is whether the anomaly-induced scale symmetry breaking is fundamental (QCD) or emergent (nuclear correlations). 
\section{Equation of State for Compact Stars}\label{compact-stars}
We now turn to the application of the $Gn$EFT formalism to the structure of massive neutron stars based on the standard TOV equation. We shall focus on the EoS of the baryonic matter, leaving out such basic issues as corrections to gravity, dark matters etc. Unless otherwise stated the role of leptons --- electrons, muons, neutrinos etc --- is included in the EOS. The results we present here are not new (as summarized in \cite{MR-review}), but  their implications and impacts on nuclear dynamics  offer a paradigm change in nuclear physics.

We will argue that what we developed above applies to massive compact stars~\cite{MR-review}.

Up to this point, we have not fixed the topology change density, apart from that it should be in the range connected to the possible ``hadron-quark transition."  To go to higher densities beyond $n_0$ in $Gn$EFT for compact stars, what  we need  is to fix the density at which the topology change takes place. It is given neither by reliable theory nor by terrestrial experiments. It turns out that the available astrophysics phenomenology does provide the range  $2\lsim n_{1/2}/n_0 < 4$~\cite{MR-review}. Here, we pick $n_{1/2}=2.5n_0$ for illustration.

Up to $n_{1/2}$, the same EoS that works well at $n_0$ is assumed to hold.  In fact available experimental data support this assumption. It is what comes at $n_{1/2}$ due to the topology change that plays the crucial role for the properties of compact stars. Among the various items listed in \cite{MR-review}, the most prominent are (a) the cusp in the symmetry energy $E_{sym}$ at $n_{1/2}$, (b) the VM with $m_\rho\to 0$ at $n_{vm}\gsim 25 n_0$, (c) the approach to the DLFP --- at or close to the putative IRFP --- at $n_{dl}\sim n_{vm}$  and (d) the hadron-quark continuity up to $n_{vm}$. The cusp in $E_{sym}$ leads  to the suppression of the $\rho$ tensor force, the most important for the symmetry energy, and triggers the effect (b). It effectively makes the EoS transit from soft-to-hard  in the EoS, thus accounting for massive $\gsim 2 M_\odot$ stars. The effects (b) and (c) --- together with the $\omega$ coupling to nucleons --- make the effective mass of the confined half-skyrmions go proportional to the dilaton decay constant $f_\chi \sim m_0$ which is independent of density. Thus {\it $f_\chi$ depends little on density in the half-skyrmion phase.}

The star properties obtained in $Gn$EFT are found to be generally consistent with presently available observations~\cite{MR-review}. There are none that are at odds with the observations -- both terrestrial and astrophysical -- within the range of error  bars quoted. For  $n_{1/2}=2.5 n_0$,  the maximum star mass is found to be $M_{max}\sim 2.0 M_\odot$ with the central density $\sim 5n_0$. For a neutron star with mass $1.4M_{\odot}$, currently highly topical in connection with gravity-wave data,  we obtain the dimensionless tidal deformability $\Lambda_{1.4} \approx 650$ and the radius $R_{1.4} \approx 12.8$~km.

However our theory makes two highly distinctive predictions that are basically different from  all  other ``standard" EFTs available in the literature. And that is in the sound velocity (SV) of stars $v_s$ and its impact on the structure of the core of massive stars, both of which are highly controversial in the community and await verdicts from experimenters.
\subsection{Emergence of pseudo-conformality}
As noted above in connection with the Gamow-Teller coupling constant in nuclei, there is scale symmetry that is hidden in nuclear Gamow-Teller matrix elements but  becomes ``visible" at high density. Here we will argue that the same symmetry manifests in the sound velocity of massive stars $v_s^2/c^2\approx 1/3$. This resembles the conformal sound velocity $v^2_{conf:s}/c^2=1/3$ expected at asymptotic density in QCD. We have suggested that what we have is not genuinely conformal because the energy momentum tensor is not traceless, i.e., $\la\theta_\mu^\mu\ra\neq  0$. This is because the compressed matter is some $\epsilon$ distance away from the IR fixed point, e.g., $m_\chi\propto \epsilon \neq 0$. However it comes out that in the range of density relevant to compact stars, $\la\theta_\mu^\mu\ra$ is {\it albeit approximately} density-independent
\be
 \frac{\del}{\del n}\la\theta_\mu^\mu\ra\approx  0.
 \ee
 One can see this in the mean-field approximation of $\psi d$HLS  (or in the large $N_c$ and $\bar{N}$ limit or Landau Fermi-liquid fixed point approximation of $Gn$EFT). The trace of the energy momentum tensor comes out to be
 \be
\la\theta^\mu_\mu\ra=4V(\la\chi\ra) -\la\chi\ra\frac{\del V(\chi)}{\del\chi}|_{\chi=\la\chi\ra}.
\ee
For simplicity we have taken the LOSS approximation, so the dilaton potential $V(\chi)$ contains all the conformal anomaly effects including quark mass terms.\footnote{We will comment below how the LOSS approximation could be in tension as in the case of quenched $g_A$.}
\subsection{Sound velocity\footnote{This and next subsections are based on results to be published elsewhere~\cite{core}.}}
Now we recall that approaching the IR fixed-point, $\la\chi\ra\sim f_\chi \neq 0$ and with the parity-doubling $f_\chi \propto m_0$ which is independent of density.  Thus
\be
\frac{\partial}{\partial n}\la \theta_\mu^\mu \ra = \frac{\partial \epsilon(n)}{\partial n}(1-3v_s^2)=0
\ee
where $v_s^2=\frac{\del P(n)}{\del n} (\frac{\del\epsilon}{\del n})^{-1}$ is the sound velocity and $\epsilon$ and $P$ are respectively the energy density and the pressure. We assume that there is no Lee-Wick-type anomalous nuclear state at the  density involved, so $\frac{\partial \epsilon(n)}{\partial n}\neq 0$ which is consistent with Nature. Therefore we have
\be
v_{pc:s}^2/c^2\approx 1/3.\label{pcs}
\ee 
We call this pseudo-conformal (PC) sound velocity.

\begin{figure}[htbp]
\begin{center}
\includegraphics[width=0.45\textwidth]{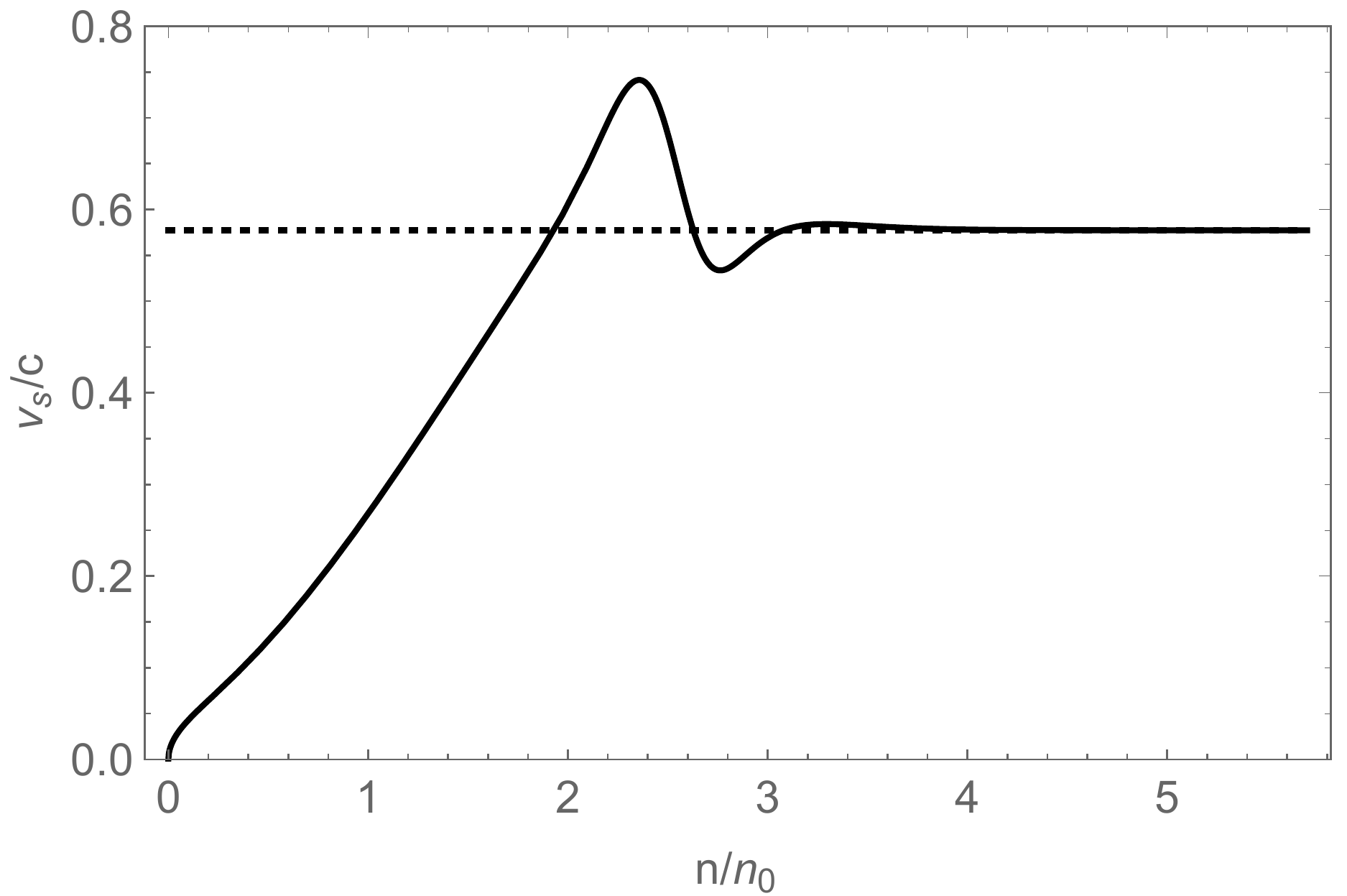}
\end{center}
\vskip -.5cm
\caption{Density dependence of the sound velocity $v_s$  in neutron matter for $n_{1/2}=2.5 n_0$. The strength of the ``spike" at $\sim n_{1/2}$ depends on $n_{1/2}$ and could exceed the causality limit for $n_{1/2} \gsim 4n_0$.}
\label{pcvs}
\end{figure}

The result (\ref{pcs}) was confirmed in the $V_{lowk}$ RG going beyond the Fermi-liquid fixed point with $1/\bar{N}$ corrections included~\cite{PKLMR,MR-review}. The result $v_{pc:s}$ obtained in the $V_{lowk}$ RG with $n_{1/2}$ set at $2.5n_0$ is given in Fig~\ref{pcvs}.

What this result shows is that the pseudo-conformality sets in slightly above $n_{1/2}$. At the point of topology change (a.k.a hadron-quark crossover), there is strong fluctuation that leads to a spike of the sound velocity (which approaches the causality limit $v_s=c$ if $n_{1/2}$ is taken at $\sim 4 n_0$~\cite{MR-review}) followed by rapid convergence to the pseudo-conformal velocity (PCV).  In some models that incorporate certain correlations such as quarkyonic structure~\cite{quarkyonic}, there appear stronger fluctuations around the density regime where microscopic quark degrees of freedom are to figure. In our description, those microscopic details are captured on the average as macroscopic dilaton property. We suggest this as a coarse-grained description of the cusp structure in the symmetry energy figuring in the spike in the presence of emerging  scale symmetry.

\subsection{Core of massive stars}
\begin{figure}[htbp]
\begin{center}
\includegraphics[width=0.45\textwidth]{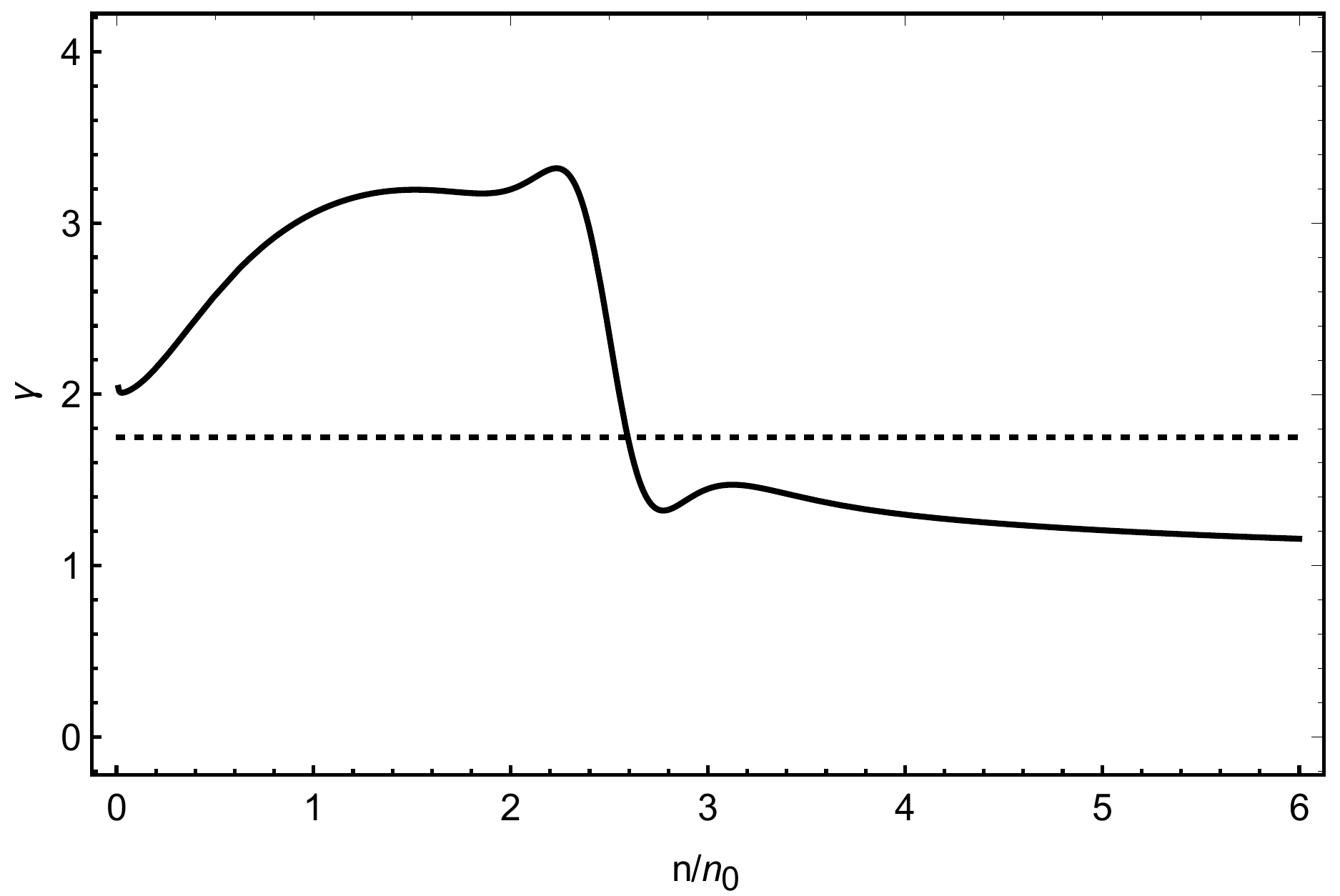}
\end{center}
\vskip -.5cm
\caption{Density dependence of  the polytropic index $\gamma= {d\ln P}/{d\ln \epsilon}$ in neutron matter for $n_{1/2}=2.5n_0$}
\label{gamma}
\end{figure}  

This prediction (\ref{pcs}) on the PCV can be confronted with a recent analysis that combines astrophysical observations and model independent theoretical {\it ab initio} calculations~\cite{evidence}. Based on the observation that, in the core of the maximally massive stars, $v_s$ approaches the conformal limit $v_s/c \to 1/\sqrt{3}$ and the polytropic index takes the value $\gamma\equiv d({\rm ln}P)/d({\rm ln}\epsilon) < 1.75$ --- the value close to the minimal one obtained in hadronic models --- Annala et al.~\cite{evidence} arrive at the conclusion that the core of the massive stars is populated by  ``deconfined" quarks.
 
It is perhaps oversimplified but surprising that the predicted pseudo-conformal speed (\ref{pcs}) sets in {\it precociously} at $\sim 3n_0$ and stays constant in the interior of the star.  Nonetheless our description does qualitatively jive with the observation of \cite{evidence}.  As already mentioned, microscopic descriptions such as the quarkyonic model~\cite{quarkyonic} typically exhibit more complex structures at the putative hadron-quark transition density. We think the simpler structure in our description is due to the suppression of higher-order $1/\bar{N}$ terms in the half-skyrmion phase. The global structure should however be robust. Similarly in Fig.~\ref{gamma}, one sees the polytropic index $\gamma$  drops, again rapidly, below 1.75 at $\sim 3n_0$ and approaches 1 at $n\gsim 6n_0$.

\begin{figure}[htbp]
\begin{center}
\vskip 0.3cm
\includegraphics[width=0.4\textwidth]{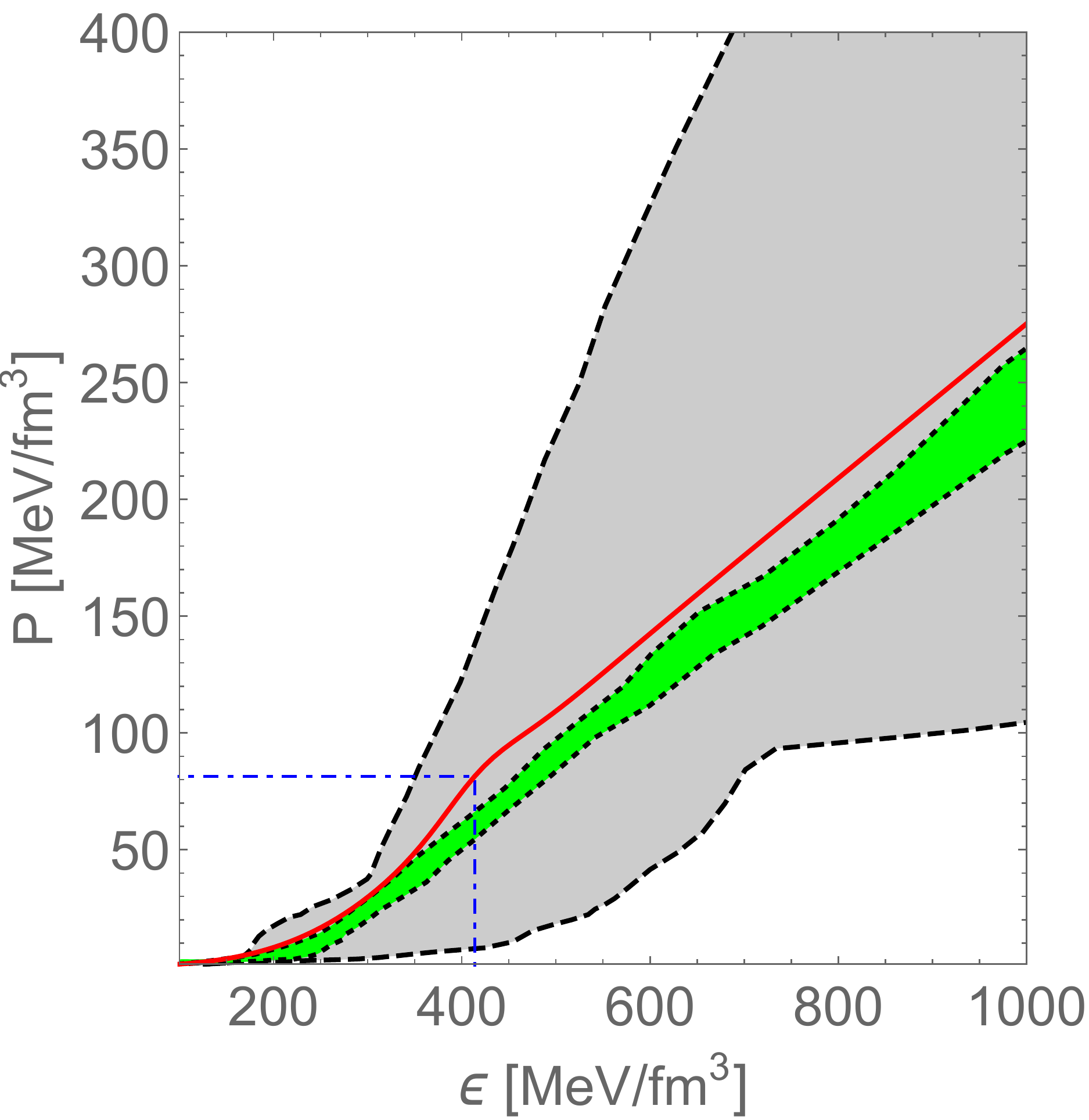}
\end{center}
\vskip -0.5cm
\caption{Comparison of $(P/\epsilon)$  between the PCM  velocity and the band generated with the
SV (sound velocity) interpolation method used in~\cite{evidence}. The gray band is  from the causality and the green band from the conformality. The red line is the PCM prediction. The dash-dotted line indicates the location of the topology change.}
\label{EoS}
\end{figure}

Finally --- and most importantly --- we compare in Fig.~\ref{EoS} our prediction for $P/\epsilon$ with the conformality band obtained by the SV interpolation method ~\cite{evidence}. We see that it is close to, and parallel with, the conformality band, but most significantly, it lies outside of this band.

The predicted results of $Gn$EFT as a whole resemble  the ``deconfined" quark structure of \cite{evidence}. There are, however, basic differences between the two. First of all,  in our theory, conformality is broken, though perhaps only slightly at high density, in the system. This could be related to the deviation of $g_A^L$ from the experimental value of the quenching in $^{100}$Sn observed~\cite{gA}. There can also be fluctuations around $v_{pcs}^2/c^2=1/3$ coming from the effects by the anomalous dimension $\beta^\prime$. This effect can be seen in Fig.~{\ref{EoS} where the PCM prediction deviates only slightly from the ``would -be" conformal band. Most important of all, the confined half-skyrmion fermion in the half-skyrmion phase is {\it not} deconfined. It is a quasiparticle of fractional baryon charge, neither purely baryonic nor purely quarkonic. In fact it can be anyonic lying on a (2+1) dimensional sheet~\cite{D}. What it represents is a manifestation of an emergent scale symmetry pervading at low density as in $g_A^L$ and  in $g_A^{DL}$ at high density in the vicinity of DLFP a.k.a IR fixed-point. We suggest to classify the precocious pseudo-conformal sound velocity in the same class of emerging scale symmetry in action in nuclear processes. In fact it is in line with how conformal symmetry permeates from the unitarity limit in light nuclei~\cite{bira-unitality} to the symmetry energy near $n_0$~\cite{lattimer-unitarity} and more.

\section{Tensions}\label{tension}
There are a few cases of tension between the predictions in $Gn$EFT  and  presently available astrophysical observables. At present none of  the tensions are  fatal but some of them are on the verge of being one. We discuss a few here and suggest possible solutions.
\subsection{Going from $R_{\rm skin}^{208}$ in $^{208}$Pb to massive compact stars?} 
The answer to the question posed above is negative in the theory formulated in this review. We discuss this as a possible tension of our approach with the structure of massive compact stars.

It has been a current lore in nuclear theory largely accepted by workers in the field that  ``precision data" in finite nuclei and nuclear matter should give {\it constraints} on the EoS for massive neutron stars. This lore  is highlighted by the recent astute and elegant attempt to have the neutron skin thickness of $^{208}$Pb, specifically the updated PREX-II, provide a stringent laboratory constraint on the density dependence of $E_{sym}$ and make statement on the neutron star matter involving much higher density than $n_0$~\cite{PREX-II}. The most interesting quantity obtained in \cite{PREX-II} is the derivative of the symmetry energy at a density in the vicinity of  $n_0$ inferred from  $R_{\rm skin}^{208}$ and is found to give
$L=3n_0 \frac{\del}{\del n} E_{sym} (n)|_{n=n_0}=(109.56\pm 34.4)\ {\rm MeV}$.
Using this data as the first ``rung in a density ladder" to go to compact-star properties, it is argued that the EoS is ``stiff" at the densities relevant {\it for atomic nuclei}. 

Indeed this $L$ value is considerably higher than what's obtained in $Gn$EFT which comes out to be $L\approx 49$ MeV~\cite{PKLMR}.  Does this raise a tension? No, we do not think so. In fact our theory gives $E_{sym}(2n_0)\approx 56$ MeV which is compatible with a presently available empirical information $(46.9\pm 12.8)$ MeV~\cite{BAL}. It is the topology change involving {\it no phase transition} that admits a ``soft" EoS below $n_{1/2}$ and ``hard" EoS above. As discussed below, the cusp structure, soft going to $n_{1/2}$ and hard going above $n_{1/2}$,  could lower  $\Lambda_{1.4}$ below $\sim 650$ Mev with $R_{1.4} < 13$ km with $n_{1/2} \gsim   2 n_0$. No phase transition(s) mentioned in \cite{PREX-II} is (are) required. This observation contraries the lore accepted in some nuclear circles that what takes place at massive star density {\it need} be constrained by what takes place at the extensively studied regime at normal matter density.  

\subsection{Confronting gravity wave}

\subsubsection{Tidal polarizability}
As noted before, the predicted upper bound for the tidal polarizability $\Lambda_{1.4}$ is $\sim 650$ with $R_{1.4}\approx 12.8$ km and $n_{cent}\approx 2.3n_0$ in $Gn$EFT. This is consistent with the present situation. However if  future measurements lower it, say, to $\sim 400$ or lower as some standard ChEFT seems to manage to bring down, then it would imply that we would have to reassess what we have done~\cite{PKLMR} -- in Fig.~\ref{EsymVlowk} -- in accounting for higher-order $1/\bar{N}$ corrections in the $V_{lowk}$ RG treatments in approaching $n_{1/2}$ from below. As shown in Fig.~\ref{Esym-cusp}, the symmetry energy at the mean field level can become {\it more attractive} as density increases from slightly below to $n_{1/2}$ in the cusp of $E_{sym}$. The central density relevant to $\Lambda_{1.4}$ is located below  $n_{1/2}$ -- i.e., ``soft" regime -- and a fine-tuning which is avoided in the treatment may be needed.   But unlike in the case of \cite{PREX-II} where the cusp-type changeover is absent, it poses no insurmountable obstacle without phase transitions.
\subsubsection{Maximum star mass $M_{max}$}
In the present approach, the range of density for the topology change allowed is $2\lsim n_{1/2}/n_0\lsim 4$. The upper bound of $n_{1/2}$, 4$n_0$, supports the maximum star mass $M_{\rm max}\simeq 2.3 M_\odot$. A somewhat higher $M_{\rm max}$ could be arrived at  by increasing $n_{1/2}$  beyond the upper bound, but  a mass as high as $M_{\rm max} \gsim 2.5 M_{\odot}$ would put our theory in tension with the pressure measured in heavy-ion collisions as noted  in \cite{PKLMR,MR-review}.  It has recently been suggested that within a $2\sigma$ confidence level the  maximum mass is $M_{\rm max}=2.210^{+0.116}_{-0.123} M_\odot$ and  the GW190814 object of $M_{\rm max}\gsim 2.5 M_\odot$ could be a black hole~\cite{Mmax}. This would lift the tension in our approach. It seems very difficult, if not impossible, within the present framework  to accommodate such a  star of mass $\gsim 2.5 M_\odot$ without a drastic revamping in taking into account $1/\bar{N}$ corrections. 
\subsection{Duality between HLS and QCD gluons}
The cusp structure of $E_{sym}$ in the skyrmion crystal simulation results from the fractionalization of the skyrmion with baryon charge $B=1$ into 2 half-skyrmions with baryon charge $B=1/2$. The change-over from skyrmions to half-skyrmions involving no Ginzburg-Landau-Wilsonian phase transitions is  a pseudogap-like phenomenon, highly exotic in nuclear physics.  In fact there can be a variety of interesting fractionalized quasiparticles, such as $1/n$ with $n=3, 5 ...$, in the skyrmion structure~\cite{canfora} or  even more intriguing sheet structures in the half-skyrmion phase~\cite{lasagne}. Also discussed in the theory literature  are  possible domain walls with deconfined quasiparticles associated with the $\eta^\prime$ ring singularity involving hidden symmetries as mentioned in Sec.~\ref{hiddensymmetries}~\cite{karasik,karasik2,kitano}. These phenomena could play an important role at high densities relevant to compact stars~\cite{D}.  Some of the topological structure imported into $Gn$EFT may be invalidated by them.  

One of the most startling novel arguments raised in the current development of hidden symmetries in strong interactions is that the hidden symmetries of the sort treated in this review, local gauge and perhaps also scale, are ``indispensable" for accessing phase transitions, be they chiral or Higgs-topology or deconfinement~\cite{karasik,karasik2,kanetal,kitano}.
\section{Further remarks} 
We have suggested that quark-like degrees of freedom, if observed in the interior of massive neutron stars, can be interpreted as confined quasi-particles of fractional baryon charges in consistency with hadron-quark continuity. Such fractionally-charged objects seem inevitable by topology at high densities~\cite{D}. The mechanism in action is the emergence of conformal (or scale) symmetry, coming not necessarily from QCD proper, but from strongly-correlated dynamical nuclear interactions, which could permeate, either hidden or exposed, in baryonic matter from low density to high density. In this scheme, true deconfinement is to set in as mentioned above at much higher densities, say, $\gsim 25 n_0$,  than relevant to compact stars when the VM fixed point and/or DLFP are reached, possibly with the phase transition from a Higgs mode to a topological mode~\cite{kanetal}.

Together with the multitude of puzzling observations, we are inevitably led to the conclusion that the densest compressed matter stable against collapse to black holes, massive neutron stars, requires a lot more than what has been debated in the literature and presents a totally uncharted domain of research.

\subsection*{Acknowledgments}
We are grateful for helpful correspondences from Jim Lattimer, Jouni Suhonen and Aleksi Vuorinen on their recent publications. The work of YLM was supported in part by the National Science Foundation of China (NSFC) under Grant No. 11875147 and 11475071.

\end{document}